\documentclass[preprint,showpacs,preprintnumbers,amsmath,amssymb]{revtex4}

\usepackage{graphicx}
\usepackage{dcolumn}
\usepackage{bm}

\newcommand{\bsym}[1]{\mbox{\boldmath$#1$}}

\begin{document}


\title{Evolution of Magnetic Fields around a Kerr Black Hole}

\author{Li-Xin Li\footnote{Chandra Fellow}}
  \affiliation{Harvard-Smithsonian Center for Astrophysics, Cambridge, 
     MA 02138}

\date{May 23, 2002}

\begin{abstract}
The evolution of magnetic fields frozen to a perfectly conducting 
plasma fluid around a Kerr black hole is investigated. We focus on 
the plunging region between the black hole horizon and the marginally 
stable circular orbit in the equatorial plane, where the centrifugal 
force is unable to stably balance the gravitational force. Adopting 
the kinematic approximation where the dynamical effects of magnetic 
fields on the fluid motion are ignored, we exactly solve 
Maxwell's equations with the assumptions that the geodesic motion 
of the fluid is stationary and axisymmetric, the magnetic field 
has only radial and azimuthal components and depends only on time 
and radial coordinates. We show that the stationary state of the 
magnetic field in the plunging region is uniquely determined by the 
boundary conditions at the marginally stable circular orbit. If the 
magnetic field at the marginally stable circular orbit is in 
a stationary state, the magnetic field in the plunging region will 
quickly settle into a stationary state if it is not so initially, 
in a time determined by the dynamical time scale in the plunging 
region. The radial component of the magnetic field 
at the marginally stable circular orbit is more important than the 
toroidal component in determining the structure and evolution of the 
magnetic field in the plunging region. Even if at the marginally 
stable circular orbit the toroidal component is zero, in the 
plunging region a toroidal component is quickly generated from the 
radial component by the shear motion of the fluid. Finally, we 
discuss the dynamical effects of magnetic fields on the motion of
the fluid in the plunging region. We show 
that the dynamical effects of magnetic fields are unimportant in 
the plunging region if they are negligible on the marginally stable 
circular orbit. This supports the ``no-torque inner boundary 
condition'' of thin disks, contrary to the claim in the recent 
literature.
\end{abstract}


\maketitle

\section{Introduction}
\label{sec1}

It is widely believed that accretion disks around Kerr black holes exist
in many astrophysical environments, ranging from active galactic nuclei to
some stellar binary systems \cite{fra02,kat98}. People usually assume that 
the inner boundary of a thin Keplerian disk around a Kerr black hole
is located at the marginally stable circular orbit, inside which the 
centrifugal force is unable to stably balance the gravity of the central 
black hole \cite{nov73,pag74}. In the disk region, particles presumably 
move on nearly circular orbits with a very small inward radial velocity 
superposed on the circular motion, the gravity of the central 
black hole is approximately balanced by the centrifugal force. As disk
particles reach the marginally stable circular orbit, the gravity of the 
central black hole becomes to dominate over the centrifugal force and the 
particles begin to nearly free-fall inwardly. The motion of fluid particles 
in the plunging region quickly becomes supersonic then the particles loose
causal contact with the disk, as a result the torque at the inner boundary 
of the disk is approximately zero \cite[and references 
therein]{muc82,abr89,pac00,arm01,afs02}. This is usually called the 
``no-torque inner boundary condition'' of thin accretion disks.

Some recent studies on the magnetohydrodynamics (MHD) of accretion disks
have challenged the ``no-torque inner boundary condition''. Magnetic fields
have been demonstrated to be the most favorable agent for the viscous
torque in an accretion disk transporting angular momentum outward
\cite[and references therein]{bal98}. By considering the evolution of 
magnetic fields in the plunging region, Krolik \cite{kro99} pointed out 
that magnetic fields can become dynamically important in the plunging 
region even though they are not so on the marginally stable circular orbit,
and argued that the plunging material might exert a torque to the disk 
at the marginally stable circular orbit. With a simplified model, Gammie 
\cite{gam99} solved Maxwell's equations in the plunging region and estimated 
the torque on the marginally stable circular orbit. He demonstrated that the 
torque can be quite large and thus the radiation efficiency of the disk can 
be significantly larger than that for a standard accretion disk where the 
torque at the inner boundary is zero. Furthermore, Agol and Krolik 
\cite{ago00} have investigated how a non-zero torque at the inner boundary 
affects the radiation efficiency of a disk.

Numerical simulations of MHD disks \cite{haw00,arm01,haw01a,haw01b,haw02} 
have greatly improved our understanding of disk accretion processes. 
These simulations show that the magneto-rotational instability effectively 
operates inside the disk and leads to accretion, though the accretion
picture is much more complicated than that assumed in the standard theory
of accretion disks. Generally, the disk accretion is non-axisymmetric and 
strongly time-dependent. It is also found that, as disk material gets into 
the plunging region, the magnetic stress at the marginally stable circular 
orbit does not vanish but smoothly extends into the plunging region 
\cite{haw01a,haw01b,haw02}, though the effect is significantly reduced
as the thickness of the disk goes down \cite{arm01,haw01a}. Furthermore,
the specific angular momentum of particles in
the plunging region does not remain constant, which implies that the
magnetic field may be dynamically important in the plunging region 
\cite{haw01a,haw01b,haw02}. All these results are fascinating and
encouraging. Unfortunately, due to the limitation in space resolution
and time integration, stationary and geometrically thin accretion
disks are not accessible to the current 2-D and 3-D simulations. So
it remains unclear how much insights we can get for stationary and 
geometrically thin accretion disks from these simulations \cite{afs02}.

Instead of small-scale and tangled magnetic fields in an accretion disk
transporting angular momentum within the disk, a large-scale and ordered 
magnetic field connecting a black hole to its disk may also exist and 
play important roles in transportation of angular momentum and energy
between the black hole and the disk \cite{bla98,li00a,li00b,li02a,li02b}.
Recent {\it XMM-Newton} observations of some Seyfert galaxies and
Galactic black hole candidates provide possible evidences for such
a magnetic connection between a black hole and its disk
\cite{wil01,li01,mil02}.

All these promote the importance of studying the evolution and the 
dynamical effects of magnetic fields around a Kerr black hole. 

In this paper, we use a simple model to study the evolution of magnetic 
fields in the plunging region around a Kerr black hole. We assume that 
around the black hole the spacetime metric is given by the Kerr metric; in 
the plunging region, which starts at the marginally stable circular orbit 
and ends at the horizon of the black hole, a stationary and axisymmetric 
plasma fluid flows inward along timelike geodesics in a small neighborhood 
of the equatorial plane. The plasma is perfectly conducting and a weak
magnetic field is frozen to the plasma. The magnetic field and the velocity 
field have two
components: radial and azimuthal. We will solve the two-dimensional 
Maxwell's equations where the magnetic field depends on two variables:
time and radius, and investigate the evolution of the magnetic field.
This model is similar to that studied by Gammie \cite{gam99}, but here we 
include the time variable. Furthermore, we ignore the back-reaction of the 
magnetic field on the motion of the plasma fluid to make the model 
self-consistent, since if the dynamical effects
of the magnetic field are important the strong electromagnetic force will
make the fluid expand in the vertical direction. The ignorance of the
back-reaction of the magnetic field will allow us to to analytically 
study the evolution of the magnetic field, but it will prevent us from
quantitatively studying the dynamical effects of the magnetic field. 
However, we believe that the essential features of the evolution of the 
magnetic field in the plunging region is not sensitive to the details of 
the dynamical effects. To check the self-consistency of the model, we 
will estimate the dynamical effects of the magnetic field by considering 
the back-reaction of the magnetic field on the fluid motion. We will also
discuss the self-consistent solutions to the coupled Maxwell and dynamical 
equations, and look for implications to the ``no-torque inner boundary 
condition''.

The paper is organized as follows: In Sec. \ref{sec2} we write down the
Maxwell's equations for an ideal MHD plasma. In Sec. \ref{sec3} we write
down the general forms of the magnetic field and the velocity field of
the plasma fluid around a Kerr black hole. In Sec. \ref{sec4} we solve 
Maxwell's equations with the approximations outlined above. In 
Sec. \ref{sec5}, with the solutions obtained in Sec. \ref{sec4}, we
study the evolution of the magnetic field in the plunging region.
In Sec. \ref{sec6}, we check if magnetic fields can become dynamically
important in the plunging region. In Sec. \ref{sec7} we draw our 
conclusions. In the Appendix we solve Maxwell's equations in the
disk region, since we want our solutions in the plunging region to
be continuously joined to the solutions in the disk region.

Throughout the paper we use the geometrized units $G = c = 1$ and the 
Boyer-Lindquist coordinates $(t,r,\theta,\phi)$ \cite{mis73,wal84},
except in the Appendix where cylindrical coordinates are used.

\section{Maxwell's equations for an ideal MHD fluid in a curved spacetime}
\label{sec2}

In a curved spacetime, Maxwell's equations are
\begin{subequations}
\begin{eqnarray}
	\nabla_a F^{ab} = - 4\pi J^b \;, \label{max1} 
        \\
	\nabla_{[a} F_{bc]} = 0\;, \hspace{0.5cm}
	\label{max2}
\end{eqnarray}
\end{subequations}
where $F_{ab} = - F_{ba}$ is the electromagnetic field tensor, $J^a$
is the current density 4-vector of electric charge. In Eqs.~(\ref{max1})
and (\ref{max2}), $\nabla_a$ denotes the covariant derivative operator
that is compatible with metric: $\nabla_a g_{bc} = 0$; the square
brackets ``[~]'' denote antisymmetrization of a tensor.	

For an ideal MHD plasma fluid whose electric resistivity is zero, the 
electric field in the comoving frame is zero, i.e.
\begin{eqnarray}
        E_a = F_{ab} u^b = 0 \;,
        \label{ezero}
\end{eqnarray}
where $u^a$ is the 4-velocity of the fluid. In other words, the magnetic 
field is frozen to the plasma fluid. Then, the electromagnetic field tensor 
$F_{ab}$ can be written as
\begin{eqnarray}
	F_{ab} = - \epsilon_{abcd} B^c u^d \;,
	\label{fab}
\end{eqnarray}
where $B^a$ is the magnetic field measured by an observer comoving with 
the fluid (i.e., having a 4-velocity $u^a$), and $\epsilon_{abcd}$ is the 
totally antisymmetric tensor of the volume element that is associated 
with the metric $g_{ab}$. By definition, the magnetic field $B^a$ satisfies
\begin{eqnarray}
	B^a u_a = 0 \;.
	\label{bua}
\end{eqnarray}
The corresponding electromagnetic stress-energy tensor is
\begin{eqnarray}
	T_{\rm EM}^{ab} = \frac{1}{4\pi} \left(F^{a}_{\;c} F^{bc}
		-\frac{1}{4} g^{ab} F_{cd}F^{cd}\right)
		= \frac{1}{4\pi} B^2 u^a u^b + \frac{1}{8\pi} 
		      B^2 g^{ab} - \frac{1}{4\pi} B^a B^b \;,
	\label{tab}
\end{eqnarray}
where $B^2 \equiv B_a B^a$\,.

In the case of MHD, the electric current density $J^a$ is unknown, but
it is only defined by Eq.~(\ref{max1}). The fundamental variables are
the electric field $E^a$ and the magnetic field $B^a$. Maxwell's
equations are then reduced to Eq.~(\ref{max2}). Since in the comoving 
frame of an ideal MHD flow $E^a = 0$ and $B^a$ is related to the 
electromagnetic tensor $F_{ab}$ by Eq.~(\ref{fab}), the dual of
Eq.~(\ref{max2}) gives \cite{lic67}
\begin{eqnarray}
	\nabla_a (u^{[a} B^{b]}) = 0 \;.
	\label{max3}
\end{eqnarray}
For an ideal MHD fluid Eqs.~(\ref{max2}) and (\ref{max3}) are equivalent.

Eq.~(\ref{max3}) can be expanded as
\begin{eqnarray}
	u^a \nabla_a B^b = B^a \nabla_a u^b - B^b \nabla_a u^a + u^b 
		\nabla_a B^a \;.
	\label{max3a}
\end{eqnarray}
The contraction of Eq.~(\ref{max3a}) with $u_b$ leads to
\begin{eqnarray}
	\nabla_a B^a - B_a a^a = 0 \;,
	\label{div}
\end{eqnarray} 
where $a^a \equiv u^b \nabla_b u^a$ is the acceleration of the fluid. In 
deriving Eq.~(\ref{div}) we have used Eq.~(\ref{bua}) and the identity 
$u_a u^a = -1$. 

Substituting Eq.~(\ref{div}) into Eq.~(\ref{max3a}), we get
\begin{eqnarray}
	u^a \nabla_a B^b = B^a \nabla_a u^b - B^b \nabla_a u^a + u^b 
		B_a a^a \;.
	\label{max3b}
\end{eqnarray}
The tensor $\nabla_a u_b$ is decomposed as \cite{haw73}
\begin{eqnarray}
	\nabla_a u_b = \frac{1}{3} \Theta h_{ab} + \sigma_{ab} - 
		\omega_{ab} -u_a a_b\;,
	\label{dab}
\end{eqnarray}
where $h_{ab} \equiv g_{ab} + u_a u_b$ is the space-projection tensor, 
$\Theta\equiv h_{ab}\nabla^a u^b = \nabla_a u^a$ is the expansion, 
$\sigma_{ab}\equiv \nabla_{(a}u_{b)} - \Theta h_{ab}/3$ is the shear tensor, 
and $\omega_{ab}\equiv - \nabla_{[a}u_{b]}$ is the vorticity tensor of the 
fluid. Here the braces ``(~)'' denote symmetrization of a tensor. It is 
easy to check that 
\begin{eqnarray}
	\sigma_{ab} u^b = \omega_{ab} u^b =0 \;, \hspace{1cm}
	\sigma_{ab} h^{ab} = \sigma_{ab} g^{ab} = 0 \;.
\end{eqnarray}
Substituting Eq.~(\ref{dab}) into Eq.~(\ref{max3b}), we obtain
\begin{eqnarray}
	u^a \nabla_a B_b = B^a (\sigma_{ab} - \omega_{ab}) - \frac{2}{3}
		\Theta B_b + B_a a^a u_b\;,
	\label{max3c}
\end{eqnarray}
which shows that the evolution of the magnetic field is governed
by the expansion, the shear, the vorticity, and the acceleration of the 
fluid.
The contraction of Eq.~(\ref{max3c}) with $B^b$ gives
\begin{eqnarray}
	\frac{1}{2} u^a\nabla_a B^2 = B^a B^b \sigma_{ab} - 
		\frac{2}{3}\Theta B^2 \;,
	\label{evol}
\end{eqnarray}
where we have used $B^a B^b \omega_{ab} = 0$ and Eq.~(\ref{bua}). 

We note that with the space-projection tensor $h_{ab}$, Eq.~(\ref{div})
can be written as 
\begin{eqnarray}
	h_{ab}\nabla^a B^b = 0 \;,
\end{eqnarray}
which says that the spatial divergence of the magnetic field is zero.

In terms of differential forms, the tensor $F_{ab}$ is a closed 2-form, 
then the Maxwell equation (\ref{max2}) can be written as $d {\bsym F} = 0$ 
\cite{wal84,sac77}. Using Stokes' theorem and Eq.~(\ref{ezero}), it can be
shown that for a perfect conducting fluid the magnetic flux threading any
2-dimensional spatial surface $S$ is unchanged as the surface moves with
the fluid
\begin{eqnarray}
        \Phi_B \equiv \int_S {\bsym F} = \mbox{constant} \;.
	\label{flux}
\end{eqnarray}
This is the mathematical formulation for the statement that magnetic field 
lines are frozen to a perfectly conducting fluid.

\section{Particle kinematics and magnetic fields around a Kerr black hole}
\label{sec3}

Now let us assume that the background spacetime is the outside of a Kerr
black hole of mass $M$ and angular momentum $Ma$, where $-M \le a \le M$. 
In Boyer-Lindquist coordinates, the Kerr metric is 
\cite{mis73,wal84}
\begin{eqnarray}
	ds^2 = -\left(1-\frac{2M r}{\Sigma}\right) dt^2 
		-\frac{4M a r}{\Sigma}\, \sin^2\theta\, dt d\phi
		+\frac{\Sigma}{\Delta}\, dr^2 +\Sigma\, d\theta^2
		+\frac{A\sin^2\theta}{\Sigma}\, d\phi^2\;,
    \label{gab}
\end{eqnarray}
where 
\begin{eqnarray}
	\Delta \equiv r^2-2M r+a^2\;, \hspace{1cm} \Sigma \equiv 
	r^2+a^2\cos^2\theta\;,
	\hspace{1cm} A \equiv (r^2+a^2)^2-\Delta a^2 \sin^2\theta\;.
	\label{del}
\end{eqnarray}
A Kerr black hole usually has two event horizons: an inner event horizon
and an outer event horizon, whose radii are given by the two roots of
$\Delta = 0$ \footnote{When $a = 0$, the Kerr black hole becomes a
Schwarzschild black hole, then the inner event horizon disappears. When
$a = \pm M$ (the case of an extreme Kerr black hole), the inner
event horizon coincides with outer event horizon.}. What is relevant in
this paper is the outer event horizon, so whenever we talk about the 
``event horizon'' we always mean the outer event horizon, whose radius 
is
\begin{eqnarray}
	r_{\rm H} = M + \left(M^2 - a^2\right)^{1/2} \;.
	\label{rh}
\end{eqnarray}

Let us define an orthonormal tetrad attached to an observer comoving 
with the frame dragging of the Kerr black hole, $\{e_0^a, e_1^a, e_2^a, 
e_3^a\}$, by
\begin{eqnarray}
	e_0^a \equiv \frac{1}{\chi}\left[\left(\frac{\partial}{\partial
		t}\right)^a + \omega \left(\frac{\partial}{\partial 
		\phi}\right)^a\right]\;, \hspace{4.3cm} 
		\nonumber \\
	e_1^a \equiv \left(\frac{\Delta}{\Sigma}\right)^{1/2} 
		\left(\frac{\partial}{\partial r}\right)^a\;, 
		\hspace{0.6cm} 
	e_2^a \equiv \frac{1}{\Sigma^{1/2}} \left(\frac{\partial}{\partial 
		\theta}\right)^a\;, \hspace{0.6cm}
	e_3^a \equiv \left(\frac{\Sigma}{A}\right)^{1/2} \frac{1}
		{\sin\theta} \left(\frac{\partial}{\partial 
		\phi}\right)^a\;,
	\label{ont2}
\end{eqnarray}
where
\begin{eqnarray}
	\chi \equiv \left(\frac{\Sigma\Delta}{A}\right)^{1/2} \;,
	\hspace{1cm}
	\omega \equiv \frac{2 M a r}{A} \;,
	\label{drag}
\end{eqnarray}
which are respectively the lapse function and the frame dragging angular
velocity. As $r\rightarrow r_{\rm H}$, we have $\chi\rightarrow 0$ and 
$\omega\rightarrow \Omega_{\rm H}$, where
\begin{eqnarray}
	\Omega_{\rm H} = \frac{a}{2 M r_{\rm H}} 
	\label{omh}
\end{eqnarray}
is the angular velocity of the event horizon. Then, the 4-velocity of 
the fluid, $u^a$, can be decomposed as
\begin{eqnarray}
	u^a = \Gamma \left(e_0^a + v_{\hat{r}} e_1^a + v_{\hat{\theta}} 
		e_2^a + v_{\hat{\phi}} e_3^a\right) \;,
	\label{ua1}
\end{eqnarray}
where $(v_{\hat{r}}, v_{\hat{\theta}}, v_{\hat{\phi}})$ are the components 
of the 3-velocity of the fluid relative to the observer comoving with the
frame dragging, and $\Gamma \equiv (1- v_{\hat{r}}^2 - 
v_{\hat{\theta}}^2 -v_{\hat{\phi}}^2)^{-1/2}$ is the Lorentz factor. 
Inserting Eq.~(\ref{ont2}) into Eq.~(\ref{ua1}), we have
\begin{eqnarray}
	u^a = \Gamma \left[\frac{1}{\chi} \left(\frac{\partial}{\partial 
		t}\right)^a + v_{\hat{r}} \left(\frac{\Delta}{\Sigma}
		\right)^{1/2} \left(\frac{\partial}{\partial r}\right)^a
		+ \frac{v_{\hat{\theta}}}{\Sigma^{1/2}} \left(\frac{
		\partial}{\partial \theta}\right)^a + \frac{\Omega}{\chi} 
		\left(\frac{\partial} {\partial \phi}\right)^a \right]\;,
	\label{ua2}
\end{eqnarray}
where 
\begin{eqnarray}
	\Omega \equiv \omega + \chi v_{\hat{\phi}} \left(\frac{\Sigma}{A}
		\right)^{1/2} \frac{1} {\sin\theta} 
	\label{angv}
\end{eqnarray}
is the angular velocity of the fluid.

The specific angular momentum of a fluid particle is
\begin{eqnarray}
	L = \left(\frac{\partial}{\partial \phi}\right)^a u_a
		= \frac{\Gamma A \sin^2\theta}{\chi\Sigma}
		(\Omega - \omega) 
		= \Gamma v_{\hat{\phi}} \left(\frac{A}{\Sigma}\right)^{1/2}
		\sin\theta \;.
	\label{sang}
\end{eqnarray}
Obviously, $L = 0$ when $\Omega = \omega$. Thus, an observer comoving with
the frame dragging of a Kerr black hole has zero angular momentum 
\cite{bar72}. The specific energy of a fluid particle is
\begin{eqnarray}
	E = - \left(\frac{\partial}{\partial t}\right)^a u_a
		= \Gamma \chi + L \omega \;.
	\label{sen}
\end{eqnarray}
In Eq.~(\ref{sen}), the term $L \omega$ represents the coupling between the
orbital angular momentum of the particle and the frame dragging of the Kerr 
black hole. When the particle moves on a geodesic, $L$ and $E$ defined 
above are conserved \cite{haw73,mis73,wal84}.

The magnetic field $B^a$, which satisfies Eq.~(\ref{bua}), can be 
decomposed as
\begin{eqnarray}
	B^a = (B_{\hat{r}} v_{\hat{r}} + B_{\hat{\theta}} v_{\hat{\theta}} 
		+ B_{\hat{\phi}} v_{\hat{\phi}}) e_0^a + B_{\hat{r}} 
		e_1^a + B_{\hat{\theta}} e_2^a + B_{\hat{\phi}} e_3^a \;,
	\label{Ba1}
\end{eqnarray}
from which we have 
\begin{eqnarray}
	B^2 \equiv B_a B^a = B_{\hat{r}}^2 + B_{\hat{\theta}}^2 + 
		B_{\hat{\phi}}^2 - (B_{\hat{r}} v_{\hat{r}} + 
		B_{\hat{\theta}} v_{\hat{\theta}} + B_{\hat{\phi}} 
		v_{\hat{\phi}})^2 \;.
	\label{bsq}
\end{eqnarray}
Inserting Eq.~(\ref{ont2}) into Eq.~(\ref{Ba1}), we have
\begin{eqnarray}
	B^a &=& \frac{1}{\chi} (B_{\hat{r}} v_{\hat{r}} + B_{\hat{\theta}} 
		v_{\hat{\theta}} +B_{\hat{\phi}} v_{\hat{\phi}}) 
		\left[\left(\frac{\partial}{\partial t}\right)^a +
		\omega \left(\frac{\partial}{\partial \phi}\right)^a\right]
		+ B_{\hat{r}} \left(\frac{\Delta}{\Sigma}
		\right)^{1/2} \left(\frac{\partial}{\partial 
		r}\right)^a \nonumber\\
		&&+ \frac{B_{\hat{\theta}}}{\Sigma^{1/2}} \left(\frac{
		\partial}{\partial \theta}\right)^a + 
		\frac{B_{\hat{\phi}}}
		{\sin\theta} \left(\frac{\Sigma}{A}\right)^{1/2}
		\left(\frac{\partial}{\partial \phi}\right)^a\;;
	\label{Ba2}
\end{eqnarray}
and, correspondingly
\begin{eqnarray}
	B_a &=& - \chi (B_{\hat{r}} v_{\hat{r}} + B_{\hat{\theta}} 
		v_{\hat{\theta}} +B_{\hat{\phi}} v_{\hat{\phi}})\, dt_a
		+ B_{\hat{r}} \left(\frac{\Sigma}{\Delta}
		\right)^{1/2}\, dr_a \nonumber\\
		&&+ B_{\hat{\theta}}\Sigma^{1/2}\, d\theta_a + 
		B_{\hat{\phi}} \left(\frac{A}{\Sigma}\right)^{1/2}
		\sin\theta\, (d\phi_a - \omega dt_a)\;.
	\label{Ba3}
\end{eqnarray}

\section{Solutions of Maxwell's equations}
\label{sec4}

The Maxwell equations that we want to solve are given by Eq.~(\ref{max3}). 
In terms of coordinate components, Eq.~(\ref{max3}) takes a very simple
form
\begin{eqnarray}
	\frac{1}{\sqrt{-g}} \frac{\partial}{\partial x^\alpha}
		\left[\sqrt{-g} \left(u^\alpha B^\beta - u^\beta
		B^\alpha\right)\right] = 0 \;,
	\label{maxeq}
\end{eqnarray}
where $x^\alpha = (t,r,\theta,\phi)$, $g$ is the determinant of the metric
tensor $g_{\alpha\beta}$, $u^\alpha \equiv u^a dx_a^\alpha$ and $B^\alpha 
\equiv B^a dx_a^\alpha$ are respectively
\begin{eqnarray}
	u^t = \frac{\Gamma}{\chi}\;, \hspace{1cm}
	u^r = \Gamma v_{\hat{r}} \left(\frac{\Delta}{\Sigma}\right)^{1/2}\;,
	\hspace{1cm}
	u^\theta = \frac{\Gamma v_{\hat{\theta}}}{\Sigma^{1/2}}\;, 
	\hspace{1cm}
	u^\phi = \frac{\Gamma}{\chi} \Omega \;,
	\label{ualp} 
\end{eqnarray}
and
\begin{eqnarray}
	B^t = \frac{1}{\chi}(B_{\hat{r}} v_{\hat{r}} + B_{\hat{\theta}} 
		v_{\hat{\theta}} +B_{\hat{\phi}} v_{\hat{\phi}}) \;, 
		\hspace{3.6cm} \nonumber\\
	B^r = B_{\hat{r}} \left(\frac{\Delta}{\Sigma}\right)^{1/2}\;,
		\hspace{1cm}
	B^\theta = \frac{B_{\hat{\theta}}}{\Sigma^{1/2}}\;, \hspace{1cm}
	B^\phi = \frac{B_{\hat{\phi}}}{\sin\theta}
		\left(\frac{\Sigma}{A}\right)^{1/2} + \omega B^t \;.
	\label{balp}
\end{eqnarray}
Because of the constraint Eq.~(\ref{bua}), among the four equations of 
(\ref{maxeq}) only three are independent.

Since the background spacetime is stationary and axisymmetric, we can look
for stationary and axisymmetric solutions with $\partial/\partial t = 
\partial/\partial \phi = 0$. However, since we are interested in the 
time evolution of magnetic fields, we will keep the $\partial/\partial 
t$ terms on magnetic fields but adopt that $\partial/\partial \phi = 0$. 
To simplify the 
problem, we further assume that in a small neighborhood of the equatorial 
plane (i.e., $|\pi/2 - \theta| \ll 1$), $u^\theta = B^\theta = 0$ (i.e., 
$v_{\hat{\theta}} = B_{\hat{\theta}} = 0$). This assumption, which has also
been used by Gammie \cite{gam99}, ensures that $\partial u^\theta/ \partial 
\theta = \partial B^\theta/ \partial \theta = 0$ on the equatorial plane. 
We emphasize that, when $\partial/\partial t\neq 0$, this assumption holds 
only if the fluid moves
geodesically, which requires that the magnetic fields are weak and their 
dynamical effects are negligible. Otherwise the electromagnetic force will
make $u^\theta$ and $B^\theta$ non-zero except exactly on the equatorial
plane. Thus, hereafter we assume that fluid particles move on timelike 
geodesics in the plunging region. This assumption will be justified latter.

Now, let us focus on solutions on the equatorial plane ($\theta = 
\pi/2$). Considering the fact that for the Kerr metric $\sqrt{-g} = 
\Sigma\sin \theta = r^2$ on the equatorial plane, Eq.~(\ref{maxeq}) is 
reduced to
\begin{eqnarray}
	\frac{\partial}{\partial t} \left[r^2 (u^t B^\beta - 
		u^\beta B^t)\right] +
		\frac{\partial}{\partial r} \left[r^2 (u^r B^\beta 
		- u^\beta B^r)\right] = 0 \;.
	\label{eq1}
\end{eqnarray}

For $\beta = r, t$, Eq.~(\ref{eq1}) gives
\begin{eqnarray}
	\frac{\partial}{\partial t}\left[r^2 (u^r B^t - u^t
		B^r)\right] = 0 \;, \hspace{1cm}
	\frac{\partial}{\partial r}\left[r^2 (u^r B^t - u^t 
		B^r)\right] = 0 \;,
	\label{eq2}
\end{eqnarray}
respectively. The solution of Eq.~(\ref{eq2}) is
\begin{eqnarray}
	u^t B^r - u^r B^t = \frac{C_0}{r^2} \;,
	\label{btr}
\end{eqnarray}
where $C_0$ is a constant. With Eq.~(\ref{flux}), it can be checked that
$2\pi C_0 \Delta\theta$ is the magnetic flux threading the 2-dimensional
surface defined by $t = {\rm constant}$, $r = {\rm constant}$, $\left\vert
\frac{\pi}{2}- \theta\right\vert \le \frac{\Delta\theta}{2}$ and $0\le
\phi < 2\pi$.

For $\beta = \theta$, Eq.~(\ref{eq1}) is automatically satisfied since
$u^\theta = B^\theta =0$ in a small neighborhood of the equatorial plane 
everywhere and all the time. 

For $\beta = \phi$, Eq.~(\ref{eq1}) gives
\begin{eqnarray}
	\frac{\partial}{\partial t} \left[r^2 (u^t B^\phi - 
		u^\phi B^t)\right] +
		\frac{\partial}{\partial r} \left[r^2 (u^r B^\phi
		- u^\phi B^r)\right] = 0 \;.
	\label{eq3}
\end{eqnarray}
From the constraint Eq.~(\ref{bua}), we have
\begin{eqnarray}
	B^\phi = - \frac{1}{u_\phi} \left(B^r u_r + B^t u_t\right) \;.
	\label{cons}
\end{eqnarray}
Substituting Eq.~(\ref{btr}) for $B^t$ into Eq.~(\ref{cons}), we obtain
\begin{eqnarray}
	B^\phi = - \frac{1}{u^r u_\phi} \left[(u^t u_t + u^r u_r) B^r 
		- \frac{C_0 u_t}{r^2}\right] \;.
	\label{bfr}
\end{eqnarray}
Now, substitute Eq.~(\ref{btr}) for $B^t$ and Eq.~(\ref{bfr}) for $B^\phi$
into Eq.~(\ref{eq3}), we obtain a first order partial differential
equation
\begin{eqnarray}
	\left(u^t \frac{\partial}{\partial t} + u^r
		\frac{\partial}{\partial r}\right) \Psi(t,r) = 0 \;,
	\label{eq4}
\end{eqnarray}
where
\begin{eqnarray}
	\Psi(t,r) \equiv \frac{1}{u_\phi} (r^2 B^r + C_0 u_t) \;.
	\label{psi}
\end{eqnarray}
In deriving Eq.~(\ref{eq4}) we have used $\partial u^\alpha/ \partial t
= 0$. Eq.~(\ref{eq4}) simply says that $\Psi$ is conserved along the
worldline of a fluid particle: $u^a \nabla_a \Psi = 0$. Let us define
\begin{eqnarray}
	\tau(r) \equiv \int_{r_0}^r \frac{u^t}{u^r}\, dr \;,
	\label{tau}
\end{eqnarray}
which is the coordinate time interval spent by a fluid particle to move 
from $r = r_0$ to $r$. Then Eq.~(\ref{eq4}) can be written as
\begin{eqnarray}
	\left(\frac{\partial}{\partial t} + \frac{\partial}
		{\partial \tau}\right) \Psi(t,\tau) = 0 \;.
	\label{eq5}
\end{eqnarray}
The solution of Eq.~(\ref{eq5}) is simply 
\begin{eqnarray}
	\Psi = \Psi (t-\tau) \;,
	\label{psol}
\end{eqnarray}
i.e. $\Psi$ is a function of $t - \tau$. 

\begin{figure}
\vspace{4.5cm}
\includegraphics[width=12cm]{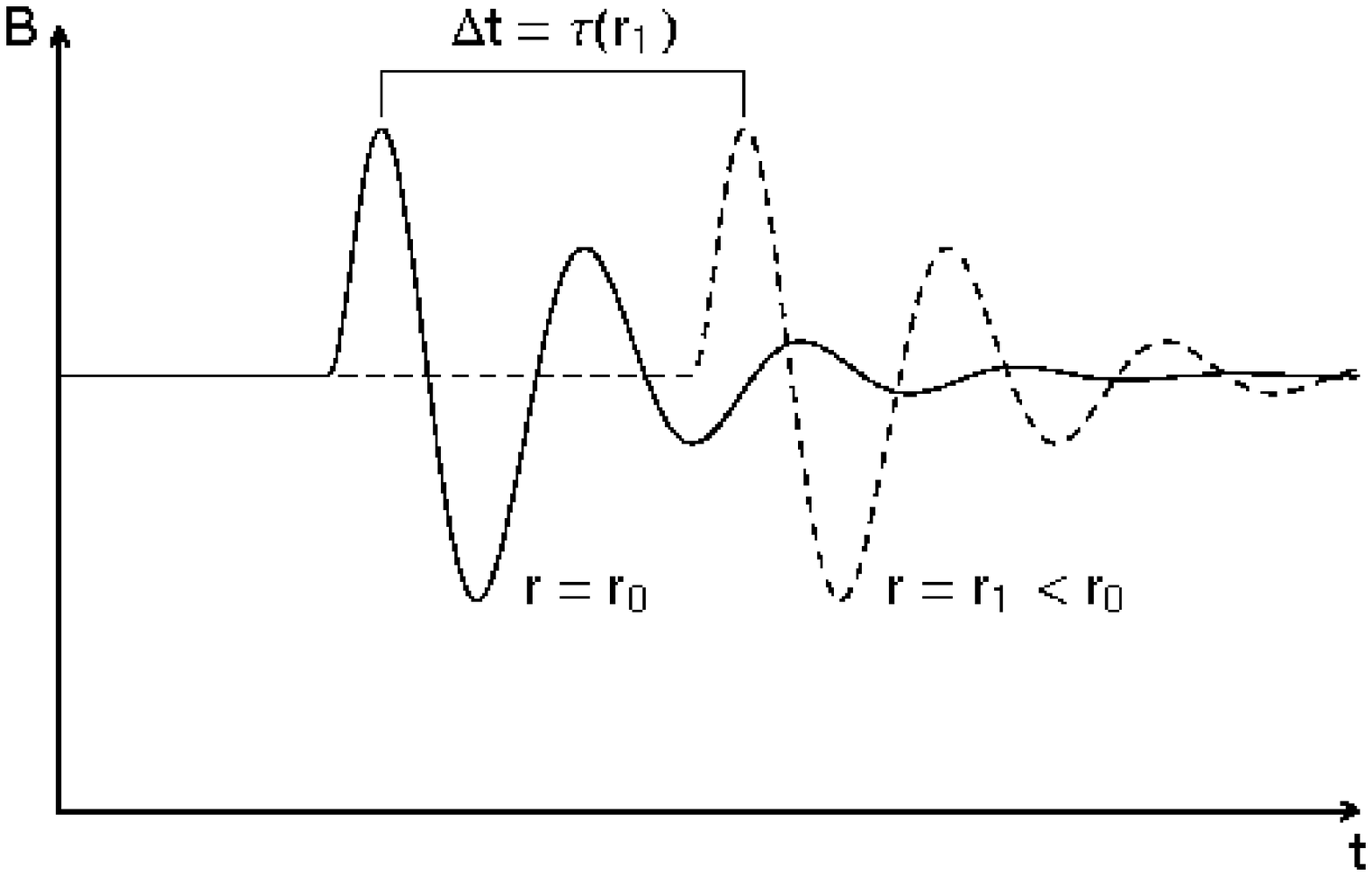}
\caption{\label{fig1} The retarded nature of the solution given by 
Eq.~(\ref{psol}). A time variation in the magnetic field at radius $r_0$
will lead to the same variation in the magnetic field at radius $r_1
< r_0$ at a later time, assuming the fluid moves toward smaller radii
(i.e. toward the central black hole). The time delay is given by the
time needed by a fluid particle moving from $r_0$ to $r_1$.}
\end{figure}

Eq.~(\ref{psol}) gives a ``retarded'' solution to Eq.~(\ref{eq5}): at 
any time $t$ the solution at radius $r$ is given by the solution at an
earlier time $t - \tau(r)$ at the radius $r_0$. Thus, a variation in 
the magnetic fields at any $r$ will propagate with the fluid motion 
(Fig.~\ref{fig1}). The solution is unique if a suitable boundary or 
initial condition is imposed. For example, if a boundary condition is 
given on $t = 0$: $\Psi (t=0, \tau) = \Psi_1 (\tau)$, then the solution 
is $\Psi(t,\tau) = \Psi_1 (\tau - t)$. In order for the solution to 
exist for a region specified by $t>0$ in the $t,\tau$-space, the boundary 
function $\Psi_1(\tau)$ must be defined on the whole $\tau$-axis: 
$-\infty < \tau < \infty$. Similarly, if a boundary condition is given 
on $\tau = 0$ (i.e., $ r = r_0$): $\Psi (t, \tau=0) = \Psi_2 (t)$, then
the solution is $\Psi(t,\tau) = \Psi_2 (t -\tau)$. In this case, 
in order for the solution to exist for a region specified by $\tau >0$ 
in the $t,\tau$-space, the boundary function $\Psi_2(t)$ must be defined 
on the whole $t$-axis: $-\infty < t < \infty$. 

\begin{figure}
\vspace{2cm}
\includegraphics[width=12cm]{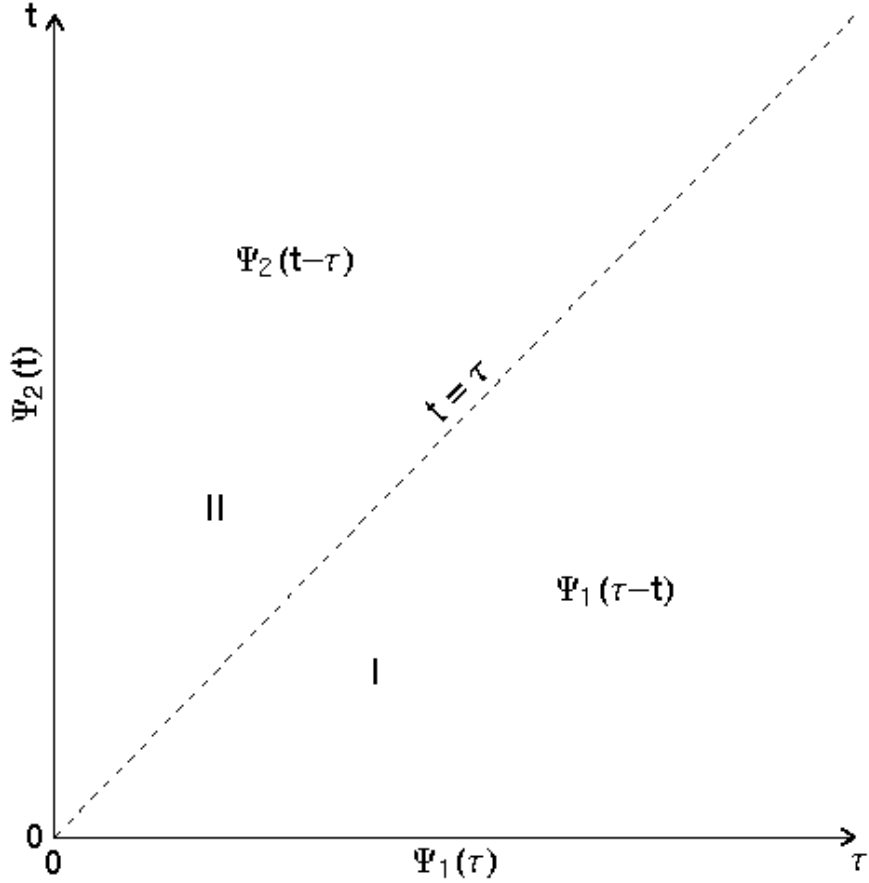}
\caption{\label{fig2} A sketch of the solution given by Eq.~(\ref{phsol}), 
corresponding to the boundary conditions given by Eq.~(\ref{bond}). The 
parameter $\tau$ is defined by Eq.~(\ref{tau}), which is the coordinate 
time needed by a particle moving from radius $r_0$ to radius $r$. So, 
$\tau = 0$ corresponds to $r = r_0$,  $\tau = \infty$ corresponds to the 
black hole horizon $r = r_{\rm H}$. The coordinate time is represented by 
$t$. In this diagram, the worldlines of fluid particles are represented by 
straight lines $t = \tau + {\rm constant}$. If the boundary conditions 
are imposed on the axes as shown in the diagram [Eq. (\ref{bond})], the 
solution in the region between the $\tau$-axis and the dashed line $t = 
\tau$ (i.e., region I) is determined
by the boundary condition on the $\tau$-axis,  the solution in the region 
between the $t$-axis and the dashed line (i.e., region II) is determined
by the boundary condition on the $t$-axis.}
\end{figure}

We can also specify the boundary condition in another way
\begin{eqnarray}
	\Psi(t = 0,\tau \ge 0) = \Psi_1 (\tau) \;, \hspace{1cm} 
	\Psi(t \ge 0, \tau = 0) = \Psi_2 (t) \;.
	\label{bond}
\end{eqnarray}
Then, the solution of Eq.~(\ref{eq5}) is
\begin{eqnarray}
	\Psi = \left\{\begin{array}{ll}
		\Psi_1(\tau - t) \;, & 0 \le t \le \tau \\
		\Psi_2(t - \tau) \;,     & 0 \le \tau \le t
		\end{array}
		\right. \;.
	\label{phsol}
\end{eqnarray}
I.e., the value of $\Psi$ in region $0 \le t \le \tau$ (region I) is 
determined by the value of $\Psi$ on the boundary $t = 0, \tau \ge 0$;
the value of $\Psi$ in region $0 \le \tau \le t$ (region II) is 
determined by the value of $\Psi$ on the boundary $\tau = 0, t \ge 0$
(see Fig.~\ref{fig2}). In order for the solutions to be smoothly matched 
on the diagonal line $t = \tau \ge 0$ separating region I and region II, 
$\Psi_1(\tau)$ and $\Psi_2(t)$ must be smoothly matched at $t = \tau = 
0$: 
\begin{eqnarray}
	\Psi_1(0) = \Psi_2(0) \;, \hspace{1cm}
	\left.\frac{d \Psi_1(\tau)}{d\tau}\right|_{\tau = 0} 
		= - \left.\frac{d \Psi_2(t)}{d t}\right|_{t = 0}\;.
\end{eqnarray}

Given the solution of $\Psi$ in Eq.~(\ref{psol}), we can solve $B^r$ from 
Eq. (\ref{psi}), then solve $B^\phi$ from Eq.~(\ref{bfr}). The results 
are
\begin{subequations}
\begin{eqnarray}
	B^r &=& \frac{1}{r^2} (-C_0 u_t + \Psi u_\phi) \;, 
		\label{fsol1}\\
	B^\phi &=& \frac{1}{r^2 u^r} [-C_0 u_t u^\phi + (1+ u^\phi 
		u_\phi)\Psi] \;.
	\label{fsol2}
\end{eqnarray}
\end{subequations}
Using Eq.~(\ref{btr}), we obtain
\begin{eqnarray}
	B^t = \frac{1}{r^2 u^r} [-(1 + u^t u_t) C_0 + u^t u_\phi 
		\Psi] \;.
	\label{fsol2a}
\end{eqnarray}
Note that $B^t$, $B^r$, and $B^\phi$ satisfy the constraint Eq.~(\ref{cons}),
so among the three components only two are independent. 

From Eqs.~(\ref{ualp}), (\ref{balp}), (\ref{fsol1}), (\ref{fsol2}), and 
the fact that $u_t = u_a (\partial/\partial t)^a = -E$ and $u_\phi = 
u_a (\partial/\partial \phi)^a = L$, we can solve for $B_{\hat{r}}$ and 
$B_{\hat{\phi}}$
\begin{subequations}
\begin{eqnarray}
	B_{\hat{r}} &=& \frac{C_0 E + \Psi L}{r \Delta^{1/2}} \;, 
		\label{br}\\
	B_{\hat{\phi}} &=& \frac{(C_0 E + \Psi L) (\Omega - \omega 
		v_{\hat{r}}^2)\Gamma}{E \chi r v_{\hat{r}}} + 
		\frac{\Psi}{E r v_{\hat{r}}} \;,
	\label{bf}
\end{eqnarray} 
\end{subequations}
where Eqs.~(\ref{sang}) and (\ref{sen}) have been used to simplify the
expression for $B_{\hat{\phi}}$. Since we focus on the solutions on the
equatorial plane, here and hereafter we set $\Sigma = r^2$, and use
$\chi$, $\omega$, and $A$ to refer their values at $\theta = \pi/2$.
Note, in the solutions in Eqs.~(\ref{fsol1}) and (\ref{fsol2}) [or,
equivalently, Eqs.~(\ref{br}) and (\ref{bf})] all the dependence on time 
$t$ is contained in the function $\Psi$. 

Since we assume that the fluid particles move on geodesics, the specific
angular momentum $L$ and the specific energy $E$ are constants. If $\Psi$ 
is also a constant, the combination $C_0 E + \Psi L$ is a constant, 
which we denote as $C_1$. Then, Eqs.~(\ref{br}) and (\ref{bf}) become
\begin{subequations}
\begin{eqnarray}
	B_{\hat{r}} &=& \frac{C_1}{r \Delta^{1/2}} \;, \label{br1}\\
	B_{\hat{\phi}} &=& \frac{C_1 \Gamma (\Omega - \omega 
		v_{\hat{r}}^2)}{E \chi r v_{\hat{r}}} + 
		\frac{\Psi}{E r v_{\hat{r}}} \;,
	\label{bf1}
\end{eqnarray}
\end{subequations}
which are stationary solutions of Maxwell's equations.

If the black hole is a Schwarzschild black hole (i.e., the specific 
angular momentum $a = 0$), the stationary solutions are reduced to
\begin{subequations}
\begin{eqnarray}
	B_{\hat{r}} &=& \frac{C_1}{\chi r^2} \;, \label{br2}\\
	B_{\hat{\phi}} &=& \frac{C_1 L}{E r^3 v_{\hat{r}}} + 
	\frac{\Psi}{E r v_{\hat{r}}} \;,
	\label{bf2}
\end{eqnarray}
\end{subequations}
where $E = \Gamma \chi$, $L = \Gamma \chi^{-1} \Omega r^2$, and $\chi^2 = 
1 - 2M/r$. The non-stationary solutions can be obtained by replacing $C_1$
with $C_0 E + \Psi L$.

\section{Evolution of magnetic fields in the plunging region}
\label{sec5}

We are interested in the evolution of magnetic fields in the plunging
region in the equatorial plane between $r = r_{\rm ms}$, the marginally 
stable circular orbit, and $r = r_{\rm H}$, the event horizon of the 
black hole. 

For direct circular orbits (i.e., corotating with $L>0$) around a Kerr
black hole, the radius of the marginally stable orbit is given by 
\cite{bar72}
\begin{eqnarray}
    r_{\rm ms} = M \left\{3+z_2-\mathrm{sign}(a) \left[(3-z_1)
		(3+z_1+2z_2)\right]^{1/2}\right\} \;,
    \label{rms}
\end{eqnarray}
where $\mathrm{sign}(a) = 1$ ($0, -1$) if $a>0$ ($= 0, <0$), and
\begin{subequations}
\begin{eqnarray}
	z_1 &\equiv& 1 + \left(1 - \frac{a^2}{M^2}\right)^{1/3}
		\left[\left(1 + \frac{a}{M}\right)^{1/3}+ \left(1 
		- \frac{a}{M}\right)^{1/3}\right] \;,\\
	z_2 &\equiv& \left(z_1^2 + 3 \frac{a^2}{M^2}\right)^{1/2} \;.
\end{eqnarray}
\end{subequations}
The angular velocity of a particle geodesically moving on the marginally 
stable circular orbit is
\begin{eqnarray}
    \Omega_{\rm ms} = \left(\frac{M}{r_{\rm ms}^3}\right)^{1/2}\left[1 
		+ a \left(\frac{M}{r_{\rm ms}^3}\right)^{1/2}\right]^{-1} 
		\;.
    \label{wr}
\end{eqnarray}
From Eqs.~(\ref{angv}), (\ref{rms}-\ref{wr}), we can calculate the
circular velocity on the marginally stable circular orbit by
\begin{eqnarray}
	v_{\hat{\phi},{\rm ms}} = \frac{A_{\rm ms}^{1/2}}{r_{\rm ms} 
		\chi_{\rm ms}}\left(\Omega_{\rm ms} 
		-\omega_{\rm ms}\right)\;,
	\label{vfrms}
\end{eqnarray}
the corresponding Lorentz factor by $\Gamma_{\rm ms} = (1 - 
v_{\hat{\phi}, {\rm ms}}^2)^{-1/2}$, specific angular momentum $L_{\rm 
ms}$ by Eq.~(\ref{sang}), and specific energy $E_{\rm ms}$ by 
Eq.~(\ref{sen}) (setting $\theta = \pi/2$ and $r = r_{\rm ms}$). Here 
and hereafter the subscript ``ms'' represents the values at $r = 
r_{\rm ms}$.

Now let us choose the boundary radius $r_0 = r_{\rm ms}$, the conserved 
specific angular momentum and specific energy to be
\begin{eqnarray}
	L = L_{\rm ms} (1 - \delta) \;,
	\hspace{0.6cm} 
	E =  E_{\rm ms}\;,
	\label{le0}
\end{eqnarray}
where $0 < \delta \ll 1$. I.e., we keep the specific energy the same as 
that on the marginally stable circular orbit, but decrease the specific 
angular momentum by a little amount. Then, we can calculate the Lorentz 
factor at $r = r_{\rm ms}$, corresponding to the specific energy and
the specific energy specified by Eq.~(\ref{le0}), by
\begin{eqnarray}
	\Gamma_0 = \frac{1}{\chi_{\rm ms}} (E - L \omega_{\rm ms}) \;;
\end{eqnarray}
and the corresponding boundary values of $v_{\hat{\phi}}$ and $v_{\hat{r}}$ 
at $r = r_{\rm ms}$ by
\begin{eqnarray}
	v_{\hat{\phi}0} = \frac{L r_{\rm ms}}{\Gamma_0 A_{\rm ms}^{1/2}} \;,
	\hspace{1cm}
	v_{\hat{r}0} = - \left(1 - v_{\hat{\phi} 0}^2 - 
		\frac{1}{\Gamma_0^2}\right)^{1/2} \;.
	\label{vrf0}
\end{eqnarray}

From Eqs.~(\ref{sang}) and (\ref{sen}), we can calculate $\Gamma$, $\Omega$, 
$v_{\hat{\phi}}$, and $v_{\hat{r}}$ at any $r$ by
\begin{subequations}
\begin{eqnarray}
	\Gamma &=& \frac{1}{\chi} (E - L \omega) \;, \\
	\Omega &=& \omega + \frac{L \chi^2 r^2}{A (E - L\omega)} \;, \\
	v_{\hat{\phi}} &=& \frac{L \chi r}{A^{1/2} (E-L\omega)} \;, \\
	v_{\hat{r}} &=& - \left[1 - \frac{\chi^2}{(E-L\omega)^2}
		\left(1 + \frac{L^2 r^2}{A}\right)\right]^{1/2} \;,
\end{eqnarray}
\end{subequations}
where $E$ and $L$ are given by Eq.~(\ref{le0}). The parameter $\tau$ defined
by Eq.~(\ref{tau}) can then be calculated with
\begin{eqnarray}
	\tau(r) = \int_{r_{\rm ms}}^r \frac{r}{\chi v_{\hat{r}}
		\Delta^{1/2}} \, dr \;.
	\label{tau1}
\end{eqnarray}

With the above formulae at hands, we can calculate $B_{\hat{r}}$, 
$B_{\hat{\phi}}$, and $B^2 = B_{\hat{r}}^2 + B_{\hat{\phi}}^2 - (B_{\hat{r}} 
v_{\hat{r}} + B_{\hat{\phi}} v_{\hat{\phi}})^2$ at any $r$ by Eqs. 
(\ref{br}) and (\ref{bf}), giving the constant $C_0$ and the function 
$\Psi(t - \tau)$. To determine $C_0$ and $\Psi$, we need to specify the 
boundary conditions for the magnetic field. We will consider stationary 
solutions and non-stationary solutions separately.

\subsection{Stationary solutions}
\label{sec51}

It is straightforward to specify the  boundary conditions for stationary 
solutions (i.e. solutions with $\Psi = {\rm constant}$). To determine the 
solutions, we need only specify  the values of $B_{\hat{r}}$ and 
$B_{\hat{\phi}}$ at $r = r_{\rm ms}$: $B_{\hat{r}0}$ and $B_{\hat{\phi}0}$. 
Then, by Eqs.~(\ref{br1}) and (\ref{bf1}), we have
\begin{subequations}
\begin{eqnarray}
	C_1 &=& B_{\hat{r}0} \chi_{\rm ms} A_{\rm ms}^{1/2}
		\;, \label{c1} \\
	\Psi &=& B_{\hat{\phi}0} E r_{\rm ms} v_{\hat{r}0} -
		B_{\hat{r}0}\Gamma_0 A_{\rm ms}^{1/2}(\Omega_0  
		- \omega_{\rm ms}v_{\hat{r}0}^2) \;,
		\label{c3}
\end{eqnarray}
\end{subequations}
where $C_1 = C_0 E + \Psi L$, $\Omega_0$ is the angular velocity 
corresponding to $v_{\hat{\phi}0}$
\begin{eqnarray}
	\Omega_0 = \omega_{\rm ms} + \frac{v_{\hat{r}0} r_{\rm ms} 
		\chi_{\rm ms}}{A_{\rm ms}^{1/2}} \;.
\end{eqnarray} 

With the $C_1$ and $\Psi$ determined above, we can calculate $B_{\hat{r}}$, 
$B_{\hat{\phi}}$, and $B^2 = B_{\hat{r}}^2 + B_{\hat{\phi}}^2 - (B_{\hat{r}} 
v_{\hat{r}} + B_{\hat{\phi}} v_{\hat{\phi}})^2$ at any $r$ by Eqs. 
(\ref{br1}) and (\ref{bf1}). Since $B_{\hat{r}}$ and $B_{\hat{\phi}}$ 
linearly depend on $B_{\hat{r}0}$ and $B_{\hat{\phi}0}$, it is sufficient
to study the effects of $B_{\hat{r}0}$ and $B_{\hat{\phi}0}$ separately.
The results for any linear combination of $B_{\hat{r}0}$ and $B_{\hat{\phi}0}$
are simply linear superpositions of the results for $B_{\hat{r}0}$ and 
$B_{\hat{\phi}0}$ separately.

\begin{figure}
\vspace{2cm}
\includegraphics[width=15cm]{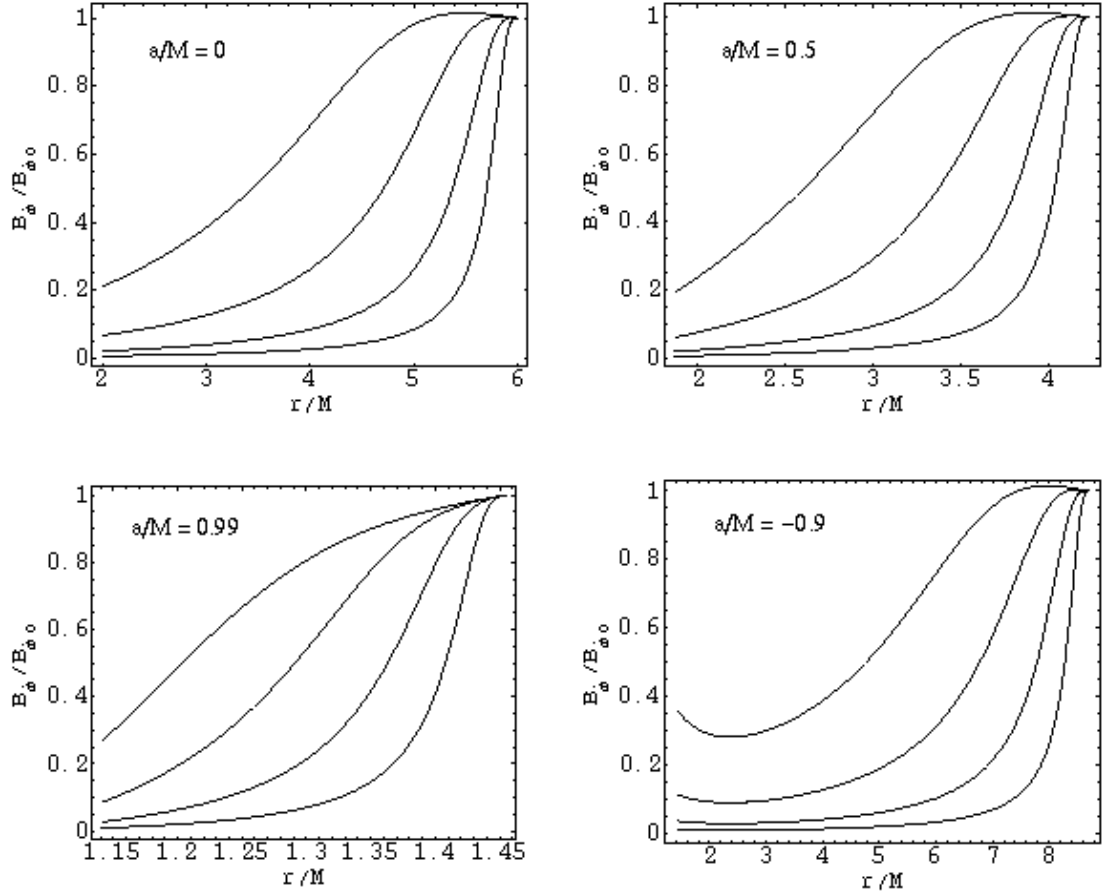}
\caption{\label{fig3} Stationary evolution of an initially toroidal magnetic
field in the plunging region around a Kerr black hole. Each panel 
corresponds to a different spinning state of the black hole, as indicated
by the dimensionless parameter $a/M$. Each curve starts from the marginally 
stable circular orbit ($r = r_{\rm ms}$, right end), ends at the horizon of 
the black hole ($r = r_{\rm H}$, left end). At $r = r_{\rm ms}$, the magnetic 
field is purely toroidal and has a value $B_{\hat{\phi}0}$, the fluid particles
have specific energy identical to that of the marginally stable circular 
orbit, and specific angular momentum smaller than that of the marginally stable 
circular orbit
by a tiny fraction $\delta$ [Eq. (\ref{le0})].  In each panel, each curve
corresponds to different values of $\delta$: $10^{-2}$, $10^{-3}$, $10^{-4}$
and $10^{-5}$ (downward). The radial component of the magnetic field is 
always zero. }
\end{figure}

In Fig.~\ref{fig3} we show the evolution results of $B_{\hat{\phi}}$
with the boundary condition $B_{\hat{\phi}} = B_{\hat{\phi}0}$ and 
$B_{\hat{r}} = 0$ at $r = r_{\rm ms}$, for different
spinning status of the black hole and different values of $\delta$ that
specify the kinetic boundary conditions of the fluid. All quantities are
scaled to the mass of the black hole so we do not need to specify the 
value of $M$. Since $B_{\hat{r}}$ does not depend on $B_{\hat{\phi}0}$,
$B_{\hat{r}}$ is always zero. Since $C_1 = 0$ and $E = {\rm constant}$, 
$B_{\hat{\phi}}$ evolves according to $B_{\hat{\phi}} \propto (r 
v_{\hat r})^{-1}$. Though in the plunging region $r$ decreases, 
$|v_{\hat r}|$ grows faster except at the neighbor of $r_{\rm ms}$.
So the evolution of $B_{\hat{\phi}}$ is dominated by the variation of
$v_{\hat r}$. Thus, as fluid particles get into the plunging region, 
$B_{\hat{\phi}}$ decreases quickly as clearly shown in Fig.~\ref{fig3}. 
This radial expansion effect is not sensitive to the spin of the black 
hole, but very sensitive to the value of $\delta$ (or, effectively, the 
initial value of $v_{\hat r}$). As $\delta$ decreases (i.e., 
$|v_{\hat{r}0}| \propto \delta^{1/2}$ decreases), the 
fluid expands more as it gets into the plunging region, so the value 
of $B_{\hat{\phi}}$ decreases more.

\begin{figure}
\vspace{2cm}
\includegraphics[width=15cm]{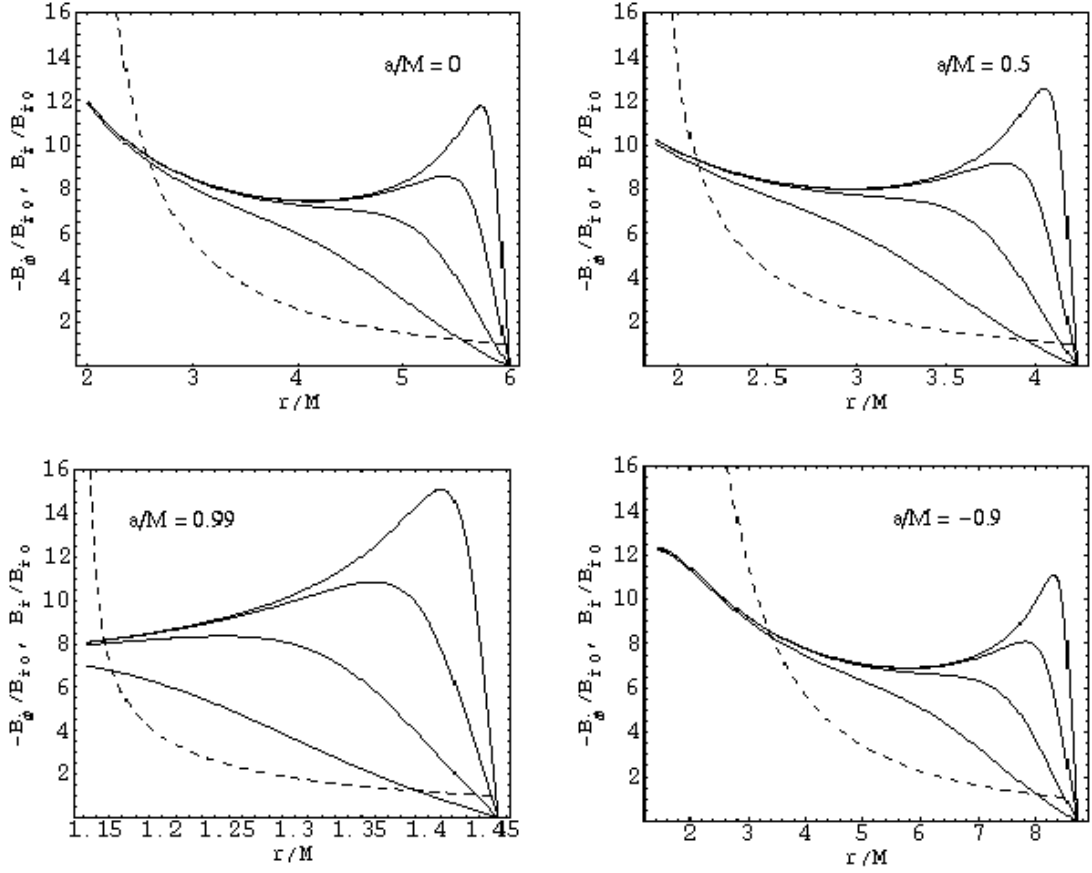}
\caption{\label{fig4} Similar to Fig.~\ref{fig3} but the magnetic field
is initially radial, i.e. $B_{\hat{r}} = B_{\hat{r}0}$ and $B_{\hat{\phi}} 
= 0$ at $r = r_{\rm ms}$. Solid lines represent the toroidal component of 
the magnetic field, dashed lines represent the radial component. Though 
initially the toroidal component of the magnetic field is zero, the 
shear motion of the fluid in the plunging region generates toroidal 
component from the radial component. The evolution of the radial component 
does not depend on the initial kinetic state of the fluid, so there is 
only one dashed line in each panel. The four solid lines representing the 
toroidal component of the magnetic field correspond to different values 
of $\delta$: $10^{-2}$, $10^{-3}$, $10^{-4}$ and $10^{-5}$ (upward).}
\end{figure}

In Fig.~\ref{fig4} we show the evolution results of $B_{\hat{r}}$ (dashed
lines) and $B_{\hat{\phi}}$ (solid lines) with the boundary condition 
$B_{\hat{r}} = B_{\hat{r}0}$ and $B_{\hat{\phi}} = 0$ at $r = r_{\rm ms}$.
Though at $r = r_{\rm ms}$ we have $B_{\hat{\phi}} = 0$,
in the plunging region $B_{\hat{\phi}}$ becomes nonzero since 
$B_{\hat{\phi}}$ depends on both $B_{\hat{r}0}$ and $B_{\hat{\phi}0}$
[Eqs. (\ref{bf1}), (\ref{c1}) and (\ref{c3})]. This is the manifestation
that the shear motion of the fluid in the plunging region generates
$B_{\hat{\phi}}$ from $B_{\hat{r}}$. The radial component of the magnetic
field, $B_{\hat{r}}$, increases gradually as the fluid enters the plunging
region, according to $B_{\hat{r}} \propto (r \Delta^{1/2})^{-1}$, and 
blows up on the black hole horizon where $\Delta = 0$. The shear motion 
of the fluid does not amplify $B_{\hat{r}}$, which echos with the fact 
that $B_{\hat{r}}$ is always zero if it is zero at $r = r_{\rm ms}$ 
(Fig.~\ref{fig3}). Since $B_{\hat{r}}$ does not depend on $v_{\hat r}$,
there is only one dashed line in each panel of Fig.~\ref{fig4}. In
comparison, the toroidal component, $B_{\hat{\phi}}$, increases more
quickly in the transition region, since the shear motion of the fluid
magnifies $B_{\hat{\phi}}$. This is more prominent for small values
of $\delta$ (i.e. small $|v_{\hat{r}0}|$), since $B_{\hat{\phi}} \propto
v_{\hat{r}}^{-1}$ [Eq. (\ref{bf1})] and $v_{\hat{r}}$ is close to 
$v_{\hat{r}0}$ as the particles just leave the marginally stable circular orbit. 
For small values of $\delta$, $B_{\hat{\phi}}$ increases sharply as the 
fluid just gets into the plunging region, then decreases a little bit 
due to the radial expansion of the fluid. Unlike $B_{\hat{r}}$, 
$B_{\hat{\phi}}$ 
is always finite on the black hole horizon. We see that, $B_{\hat{r}}$ 
and $B_{\hat{\phi}}$ evolves in very different ways. 

From Fig.~\ref{fig4} we also see that the evolution of the magnetic
field in the plunging region depends on the spin of the black hole,
though not very sensitively. Interestingly, the toroidal component and
the poloidal component depend on the spin of the black hole in an opposite
way. As the dimensionless spin parameter $a/M$ increases from zero to 
positive values, the shear amplification effect on the toroidal component
of the magnetic field increases (except for the case of $\delta = 10^{-2}$
for which the amplification effect is not prominent), while the
amplification of the radial component caused by the compression in the
azimuthal direction (i.e., the decrease in radius $r$) decreases. But,
if $a/M$ decreases from zero to negative values, the shear amplification 
effect on the toroidal component of the magnetic field decreases, while
the compression amplification of the radial component increases. The 
opposite dependence for positive and negative spins is probably due to 
the different coordinate distances from the marginally stable circular orbit to 
the black hole horizon for black holes of positive and negative spins.

In Fig.~\ref{fig5} we show the evolution of $B^2$ [defined by
Eq. (\ref{bsq})] with the same boundary conditions in Fig.~\ref{fig4}.
As the fluid just enters the plunging region (near the right ends of 
curves), $B^2$ sharply increases due to the small values of 
$|v_{\hat{r}}|$ there, which is the manifestation of amplification 
effect caused by the shear rotation of the fluid. After that, i.e., 
after the fluid obtains a large radial velocity (the dashed
lines in the figure), the amplification effect is reduced but the
expansion effect becomes prominent [see Eq. (\ref{evol})]. On the horizon 
of the black hole $B^2$ is always finite, so the boundary conditions on 
the horizon is satisfied \cite[and references therein]{tho86}.

\subsection{Non-stationary solutions}
\label{sec52}
To specify the boundary conditions for non-stationary solutions is a little 
bit complicated. As discussed earlier, to determine the solutions
in a region with $t > 0$ and $\tau > 0$ (i.e. $r_{\rm H} < r < r_{\rm 
ms}$) in the $t,\tau$-space, we need to specify the boundary conditions 
on the axes ($t = 0, \tau \ge 0$) and ($t \ge 0, \tau = 0$), i.e. 
specify $\Psi_1(\tau)$ and $\Psi_2(t)$ [see Eq.~(\ref{bond})]. 

As an example, let us assume that $\Psi_2(t) = {\rm constant}$, and 
$B_{\hat{\phi}} = 0$ on the axis ($t = 0, \tau \ge 0$). I.e., the 
solution on the boundary $\tau = 0$ is stationary for $t \ge 0$, 
and the $\hat{\phi}$-component of the magnetic field in the plunging 
region is zero at $t = 0$. Equivalently, we specify the boundary 
conditions as follows:
\begin{subequations}
\begin{eqnarray}
	{\rm On}\;\, r = r_{\rm ms},\; t \ge 0: \hspace{1.453cm}
		B_{\hat{r}} = B_{\hat{r}0},\;\; B_{\hat{\phi}} = 0 
		\;; \label{bond1}\\
	{\rm On}\;\, t = 0,\; r_{\rm H} <r < r_{\rm ms}:
		\hspace{0.5cm} B_{\hat{\phi}} = 0 \;. \hspace{2.12cm} 
		\label{bond2}
\end{eqnarray}
\end{subequations}
Then, if we apply Eqs.~(\ref{br}) and (\ref{bf}) to the boundary at
$r = r_{\rm ms}$ and $t \ge 0$, we can solve for $C_0$ and $\Psi_2$.
The results are
\begin{subequations}
\begin{eqnarray}
	C_0 &=& \frac{B_{\hat{r}0} A_{\rm ms}^{1/2}}{E}
		\left[(\Omega_0 - \omega_{\rm ms} v_{\hat{r}0}^2)
		L \Gamma_0 + \chi_{\rm ms}\right] \;, \label{c0s} \\
	\Psi_2 &=& - \Gamma_0 B_{\hat{r}0} A_{\rm ms}^{1/2}
		(\Omega_0 - \omega_{\rm ms} v_{\hat{r}0}^2) \;.
		\label{psi2}
\end{eqnarray}
\end{subequations}
Applying Eq.~(\ref{bf}) to the boundary at $t = 0$ and $r_{\rm H} < r
< r_{\rm ms}$, and substituting Eq.~(\ref{c0s}) for $C_0$, we can solve
for $\Psi_1$
\begin{eqnarray}
	\Psi_1(\tau) = \Psi_1[r(\tau)] \;, \hspace{1cm}
	\Psi_1(r) = \frac{(\Omega - \omega v_{\hat{r}}^2) \Gamma E C_0}
		{(\Omega - \omega v_{\hat{r}}^2) \Gamma L + \chi} \;,
	\label{psi1}
\end{eqnarray}
where $r(\tau)$ is given by the inverse of $\tau(r)$. The values of
$B_{\hat{r}}$ on $t = 0$ and $r_{\rm H} < r < r_{\rm ms}$ is then 
determined by Eq.~(\ref{br}).

With the $\Psi_1$ and $\Psi_2$ determined above, we can obtain $\Psi$
by Eq.~(\ref{phsol}). This, together with the $C_0$ given by 
Eq.~(\ref{c0s}), allows us to calculate $B_{\hat{r}}$ and $B_{\hat{\phi}}$
with Eqs.~(\ref{br}) and (\ref{bf}) for any radius in the plunging region
at any time $t>0$. 

It is easy to check that $\Psi_1 (\tau = 0) = \Psi_2$, so the solutions 
are continuously matched on the line $t = \tau$. Since 
\begin{eqnarray}
	\frac{d \Psi_1(\tau)}{d \tau} = \frac{d \Psi_1(r)}{d r} \,
		\frac{dr}{d\tau} = \frac{d \Psi_1(r)}{d r} \,
		\frac{\chi v_{\hat{r}} \Delta^{1/2}}{r} \;,
\end{eqnarray}
the solutions are smoothly matched on the line $t = \tau$ only in the limit
$v_{\hat{r}0} = 0$ since then we have $d \Psi_1/d\tau = 0 = - d \Psi_2 /dt$ 
at $\tau = t =0$. If $v_{\hat{r}0}$ is not zero but small ($|v_{\hat{r}0}| 
\ll 1$), the solutions are approximately smoothly matched on the line 
$t = \tau$ if $|d\Psi_1(r)/dr|$ is not large at $r = r_{\rm ms}$.

In Figs.~\ref{fig6} -- \ref{fig8} we show the results for the non-stationary 
evolution
of magnetic fields. Each figure corresponds to a different spinning state of 
the black hole. In Fig.~\ref{fig6} $a/M = 0$, in Fig.~\ref{fig7} $a/M = 0.99$,
while in Fig.~\ref{fig8} $a/M = -0.9$. The boundary conditions for the 
magnetic field are given by Eqs.~(\ref{bond1}) and (\ref{bond2}). The kinetic 
boundary conditions of the fluid are given by Eq.~(\ref{le0}), here we assume 
$\delta = 10^{-3}$. In each figure, each panel corresponds to the state of
the magnetic field at a particular moment. Especially, the first (left and
up) panel shows the initial ($t = 0$) state of the magnetic field [i.e.
Eqs. (\ref{bond1}) and (\ref{bond2})]. In each panel, the thick dashed line
shows $B_{\hat{r}}$, the thick solid line shows $B_{\hat{\phi}}$, and the
thin lines show the corresponding stationary solutions. Each curve starts
from the marginally stable circular orbit (right end) and ends on the horizon 
of the black hole (left end). 

From these figures we see that, though initially the magnetic field is 
deviated from the stationary state, it evolves to and finally saturates at the 
stationary state as time goes on. Since we hold the magnetic field on the
marginally stable circular orbit in a stationary state for $t \ge 0$ 
[Eq.~(\ref{bond1})] and the fluid moves from the marginally stable circular 
orbit toward the horizon of the black hole, the stationary state propagates 
from the marginally stable circular orbit toward the horizon of the black hole 
(i.e., from right to left in the figures). This is clear seen in 
Figs.~\ref{fig6} -- \ref{fig8}: the magnetic field at a radius closer to the 
right end of the curve gets into the stationary state earlier. In each panel, 
with a black dot we show the 
position of a particle that is at the marginally stable radius at $t = 0$,
which clearly shows the propagation of the state of the magnetic field 
with the motion of the fluid. This suggests that the stationary state of the 
magnetic field in the plunging region is uniquely determined by the boundary
conditions at the marginally stable circular orbit.

The figures also show the dependence of the evolution of the magnetic
field on the spin of the black hole. For a black hole with a negative 
$a$ a longer coordinate time is needed for most of the fluid in the 
plunging region to settle into the stationary state. This is apparently
caused by the fact that a black hole with a negative $a$ has a larger
coordinate radial distance from the marginally stable circular orbit
to the horizon.

\section{Can magnetic fields become dynamically important in the 
plunging region?}
\label{sec6}

In previous sections we have studied the evolution of magnetic fields in 
the plunging region and seen that magnetic fields can be amplified by the 
convergence (the radial magnetic field) and shear motion (the toroidal 
magnetic field) of the fluid. Thus, we can ask if magnetic fields can become 
dynamically important in the plunging region assuming that their dynamical 
effects are negligible at the marginally stable circular orbit. We try to 
answer this question in this section.

\subsection{Dynamical effects of the magnetic field}
\label{sec61}

Assuming that in the plunging region the gas pressure of the fluid is 
negligible, then the total stress-energy tensor of the fluid is
\begin{eqnarray}
	T^{ab} = \rho_{\rm m} u^a u^b + T_{\rm EM}^{ab} \;,
	\label{tab1}
\end{eqnarray}
where $\rho_{\rm m}$ is the mass-energy density of the fluid matter
measured by an observer comoving with the fluid, $T_{\rm EM}^{ab}$ is the
stress-energy tensor of the electromagnetic field as given by Eq.~(\ref{tab}).
Then, the equation of motion of the fluid is given by 
\begin{eqnarray}
	\nabla_a T^{ab} = \rho_{\rm m} u^a \nabla_a u^b + u^b
		\nabla_a (\rho_{\rm m} u^a) + \nabla_a T_{\rm EM}^{ab}
		= 0 \;.
	\label{dtab}
\end{eqnarray}

When Maxwell's equations are satisfied we have $\nabla_a T_{\rm EM}^{ab}
= - F^{ab} J_b$ \cite{mis73}. Then, by Eq.~(\ref{ezero}), we have $u_b 
\nabla_a T_{\rm EM}^{ab} = 0$ for a magnetic field frozen to the fluid. 
Therefore, the contraction of $u_b$ with 
Eq.~(\ref{dtab}) leads to the equation of continuity \footnote{When the 
gas pressure of the fluid is not negligible, the equation of continuity 
(\ref{cont1}) also holds if the motion of the fluid is locally adiabatic 
\cite{lic67}.}
\begin{eqnarray}
	\nabla_a J_{\rm m}^a = 0 \;, \hspace{1cm}
	J_{\rm m}^a \equiv \rho_{\rm m} u^a \;,
	\label{cont1}
\end{eqnarray}
where $J_{\rm m}^a$ is the flux density vector of mass. The equation of
continuity does not explicitly contain magnetic field variables.
Since $(\partial/\partial t)^a$ and $(\partial/\partial \phi)^a$ are
Killing vectors, Eq.~(\ref{dtab}) leads to the conservation of angular 
momentum and energy
\begin{eqnarray}
	\nabla_a J_{\rm L}^a = 0 \;, \hspace{1cm}
	\nabla_a J_{\rm E}^a = 0 \;,
\end{eqnarray}
where
\begin{eqnarray}
	J_{\rm L}^a \equiv \left(\frac{\partial}{\partial 
		\phi}\right)^b T_b^{\,\,a}
		= \left(\rho_{\rm m} +\frac{1}{4\pi} B^2\right) 
		L u^a - \frac{1}{4\pi} B_\phi B^a 
\end{eqnarray}
is the flux density vector of angular momentum,
\begin{eqnarray}
	J_{\rm E}^a \equiv - \left(\frac{\partial}{\partial 
		t}\right)^b T_b^{\,\,a} 
		= \left(\rho_{\rm m} +\frac{1}{4\pi} B^2\right) 
		E u^a + \frac{1}{4\pi} B_t B^a
\end{eqnarray}
is the flux density vector of energy.

Here we look for stationary and axisymmetric solutions in the plunging
region. Then, with the assumptions adopted in this paper, i.e. $\partial/
\partial t = \partial/\partial \phi = 0$, $u^\theta = B^\theta = 0$ and 
$\partial u^\theta /\partial\theta = \partial B^\theta/\partial\theta = 
0$, in the equatorial plane the equation of continuity is reduced to
\begin{eqnarray}
	\frac{d}{d r}\left(r^2 \rho_{\rm m} u^r\right) = 0 
		\;.
	\label{cont2}
\end{eqnarray}
Similarly, the equations of angular momentum and energy conservation are 
reduced to
\begin{eqnarray}
	\frac{d}{d r} \left\{r^2 \left[\left(\rho_{\rm m} +
		\frac{1}{4\pi} B^2\right) L u^r - \frac{1}{4\pi} B_\phi
		B^r\right]\right\} = 0 \;, \label{lcon1}
\end{eqnarray}
and
\begin{eqnarray}
	\frac{d}{d r} \left\{r^2 \left[\left(\rho_{\rm m} +
		\frac{1}{4\pi} B^2\right) E u^r + \frac{1}{4\pi} B_t
		B^r\right]\right\} = 0 \;. \label{econ1}
\end{eqnarray}

The solution of Eq.~(\ref{cont2}) is
\begin{eqnarray}
	\rho_{\rm m} = \frac{F_{\rm m}}{4 \pi r^2 u^r} =
		\frac{F_{\rm m}}{4 \pi \Gamma \chi A^{1/2} 
		v_{\hat{r}}} \;,
	\label{rhm}
\end{eqnarray}
where $F_{\rm m} = 4 \pi r^2 J_{\rm m}^r$ is a constant to be determined by 
the boundary conditions, which measures the mass flux across 
radius $r$. Fig.~\ref{fig9} shows the variation of the mass density with 
radius in the plunging region, assuming that fluid particles move on 
geodesics. The kinetic boundary conditions are given by Eq.~(\ref{le0}). 
In each panel (corresponding to a different spinning
state of the black hole), we show the ratio of $\rho_{\rm m}/\rho_{{\rm m}
0}$ corresponding to four different values of $\delta$: $10^{-2}$, 
$10^{-3}$, $10^{-4}$ and $10^{-5}$, where $\rho_{{\rm m}0} \equiv \rho(r = 
r_{\rm ms})$. We see that, the evolution of the mass density
of the fluid sensitively depends on the value of $\delta$ (or effectively,
the value of $|v_{\hat{r} 0}| \propto \delta^{1/2}$). While for not very
small $\delta$ (e.g., $\delta = 10^{-2}$) the variation of $\rho_{\rm m}$
is not dramatic, for very small $\delta$ (e.g., $\delta = 10^{-5}$) the
variation of $\rho_{\rm m}$ is dramatic: as the fluid gets into the
plunging region the mass density drops sharply. This is caused by the sharp
increase in the ratio $v_{\hat{r}}/v_{\hat{r} 0}$ in the plunging region
for small $\delta$.

From Eqs.~(\ref{lcon1}) and (\ref{econ1}) we see that the dynamical role 
played by the magnetic field is characterized by the following three 
dimensionless parameters
\begin{eqnarray}
	\eta_1 \equiv \frac{B^2}{4\pi \rho_{\rm m}} \;, 
		\hspace{1cm}
	\eta_2 \equiv \left\vert\frac{B_\phi B^r}{4\pi 
		\rho_{\rm m} L u^r}\right\vert \;, \hspace{1cm} 
	\eta_3 \equiv \left\vert\frac{B_t B^r}{4\pi 
		\rho_{\rm m} E u^r}\right\vert \;.
	\label{eta}
\end{eqnarray}
The parameter $\eta_1$ measures the transfer of angular momentum and 
energy from the fluid particles to the magnetic field; $\eta_2$ and
$\eta_3$ measure the transportation of angular momentum and energy from 
one part to another by the magnetic tension in the rest frame of the fluid.
To justify the assumption that the effects of the magnetic field on the
dynamics of the fluid particles are negligible, we must require that
\begin{eqnarray}
	\eta_1 \ll 1 \;, \hspace{1cm}
	\eta_2 \ll 1 \;, \hspace{1cm} 
	\eta_3 \ll 1 \;.
	\label{nc1}
\end{eqnarray} 

In Eq.~(\ref{eta}), $\rho_{\rm m}$ is given by Eq.~(\ref{rhm}), $B^2$ is 
calculated by Eq.~(\ref{bsq}) (setting $v_{\hat{\theta}} = B_{\hat{\theta}}
= 0$), $B_\phi B^r$ and $B_t B^r$ are calculated by
\begin{eqnarray}
	B_\phi B^r = \frac{\chi A}{r^3} B_{\hat{r}} B_{\hat{\phi}} \;,
	\hspace{1cm}
	B_t B^r = - \frac{\chi A^{1/2}}{r^3} B_{\hat{r}}
		\left(\chi r B_{\hat{r}} v_{\hat{r}} + A^{1/2} 
		B_{\hat{\phi}} \Omega\right) \;.
	\label{btbr}
\end{eqnarray}
Near the marginally stable circular orbit, we have $|v_{\hat{r}}/v_{\hat{\phi}}|
\ll 1$, $E \ge L \Omega$, and $B_t B^r \approx - \Omega B_\phi B^r$.
Therefore, from the definitions of $\eta_2$ and $\eta_3$, we have
\begin{eqnarray}
	\eta_3 \approx \frac{\Omega L}{E} \eta_2 
		\le \eta_2 \;,
	\label{et32}
\end{eqnarray}
at $r \approx r_{\rm ms}$. 

From Fig.~\ref{fig9} we see that for small $\delta$ the mass density
$\rho_{\rm m}$ drops quickly as the fluid enters the plunging region.
However, the evolution of $B^2$ in the plunging region sensitively depends
on the orientation of the magnetic field on the marginally stable circular orbit.
From Fig.~\ref{fig5}, the value of $B^2$ corresponding to an
initially radial magnetic field increases as the fluid enters the plunging 
region, because of the convergence of the radial magnetic field and the 
shear amplification of the toroidal magnetic field. If the initial magnetic 
field is purely toroidal, on the other hand, from Fig.~\ref{fig3} we see 
that the magnetic field keeps purely toroidal in the plunging region and 
the dilution effect arising from the drop in the mass density makes 
$B_{\hat{\phi}}$ (and thus $B^2$) decrease in the plunging region. 
Therefore, in the plunging region the variation of $\eta_1$, $\eta_2$, 
and $\eta_3$ sensitively depends on the boundary
condition of the magnetic field on the marginally stable circular orbit. We can 
imagine that, if on the marginally stable circular orbit the magnetic field is
purely radial, then in the plunging region $B^2$ increases so the
ratio $\eta_1$ also increases as $\rho_{\rm m}$ decreases. In such a case, 
$\eta_1$ may become close to or even greater than $1$ in the plunging 
region even if it is $\ll 1$ on the marginally stable circular orbit, then the 
dynamical effects of the magnetic field become important in the plunging
region. On the other hand, 
if on the marginally stable circular orbit the magnetic field is purely toroidal, 
then in the plunging region $B^2$ decreases which causes $\eta_1$ to 
decrease also if $B^2$ decreases faster than $\rho_{\rm m}$. Therefore, to 
correctly estimate the dynamical effects of magnetic fields in the plunging 
region, we must choose a sensible boundary condition for the magnetic field 
on the marginally stable circular orbit.

The magnetic field on the marginally stable circular orbit is determined
by the MHD processes in the disk region. In the disk region ($r>r_{\rm 
ms}$), particles move on nearly circular orbits with $|v_{\hat{r}}| \ll
|v_{\hat{\phi}}|$. As a disk particle moves a finite distance in the radial
direction, it has undergone an infinite number ($\gg 1$)
of turns around the central black hole. As a result, the magnetic field
lines frozen to the fluid are wound up around the center an 
infinite number of times. Therefore, in a stationary state, we expect
that the magnetic field in the disk region is predominantly toroidal. In
the Appendix we solve Maxwell's equations in the disk region and show
that in the stationary state the magnetic field in the disk is likely to 
be parallel to the velocity field: $B^\phi / B^r = u^\phi /u^r$. Thus, in 
the stationary state, on the marginally stable circular orbit the magnetic 
field does not take any orientation but the one that satisfies the following 
condition
\begin{eqnarray}
        \frac{B^\phi u^r}{B^r u^\phi} = 1 \;.
        \label{pc}
\end{eqnarray}
On the marginally stable circular orbit, Eq.~(\ref{pc}) implies that $|B_{\hat{r}}| 
\ll |B_{\hat{\phi}}|$ since $|v_{\hat{r}}| \ll |v_{\hat{\phi}}|$.

From Eqs.~(\ref{fsol1}) and (\ref{fsol2}), we have
\begin{eqnarray}
        \frac{B^\phi u^r}{B^r u^\phi} = 1 + \frac{\Psi}
                {(-C_0 u_t + \Psi u_\phi) u^\phi} \;.
        \label{bfru}
\end{eqnarray}
Then, Eq.~(\ref{pc}) is satisfied if and only if $\Psi = 0$ on the
marginally stable circular orbit. In the stationary state $\Psi$ is a constant,
so we have 
\begin{eqnarray}
        \Psi = 0
        \label{psi0}
\end{eqnarray}
throughout the plunging region. Thus, in the stationary state, 
Eq.~(\ref{pc}) holds throughout the plunging region \footnote{Occasionally
we say that the magnetic field is parallel to the velocity field (or,
equivalently, the orientation of the magnetic field follows the 
orientation of the velocity field), if Eq.~(\ref{psi0}) [or, 
equivalently, Eq.~(\ref{pc})] is satisfied. However, we point out that
this does not always mean that the corresponding electric field 
measured by an observer comoving with the frame dragging is zero, unless
the black hole is a Schwarzschild black hole. One can check that, for
the solutions in Eqs~(\ref{fsol1}) and (\ref{fsol2}), the corresponding
electric field measured by an observer comoving with the frame dragging 
[i.e., with a 4-velocity $e_0^a$ given in Eq.~(\ref{ont2})] is $E_a = 
\chi^{-1}(C_0 \omega + \Psi) d\theta_a$\,. So, when $\Psi = 0$, we have
$E_a = C_0 \chi^{-1} \omega d\theta_a$\,, which is zero only if $a = 
0$ (then the frame-dragging frequency $\omega = 0$).}.

Inserting Eqs.~(\ref{ualp}) and (\ref{balp}) into Eq.~(\ref{pc}), we
obtain
\begin{eqnarray}
        \frac{B_{\hat{r}}}{B_{\hat{\phi}}} = \frac{E r v_{\hat{r}}}
                {\Gamma A^{1/2} (\Omega - \omega v_{\hat{r}}^2)} \;.
        \label{pc1}
\end{eqnarray}

We have calculated $\eta_1$, $\eta_2$, and $\eta_3$ in the plunging region, 
assuming that they all are $\ll 1$ at $r = r_{\rm ms}$ and the boundary
condition (\ref{psi0}) is satisfied. At $r = r_{\rm ms}$ the 
toroidal magnetic field is assumed to be $B_{\hat{\phi}} = 0.05$ in units 
of $(4 \pi \rho_{\rm m})^{1/2}$, the corresponding radial magnetic field 
$B_{\hat{r}}$ is then given by Eq.~(\ref{pc1}). The parameter $\delta$ 
is taken to be $10^{-2}$, $10^{-3}$, $10^{-4}$ and $10^{-5}$, alternatively. 
The results are shown in Figs.~\ref{fig10} and \ref{fig11}. From these 
figures we see that, in the plunging region the evolution of $\eta_1$, 
$\eta_2$ and $\eta_3$ sensitively depends on the value of $\delta$. For
very small $\delta$, $\eta_1$, $\eta_2$ and $\eta_3$ quickly decrease as 
the fluid gets into the plunging region. This is caused by the fact that 
a very small $\delta$ (i.e., a very small $|v_{\hat{r}0}| \propto 
\delta^{1/2}$) corresponds to a very small $|B_{\hat{r}0}|$ according to
the boundary condition (\ref{pc}), while the magnetic field in the 
plunging region is predominantly determined by $|B_{\hat{r}0}|$ instead
of $|B_{\hat{\phi}0}|$ (see Figs.~\ref{fig3} and \ref{fig4}). For a moderate
$\delta$ (e.g., $\delta = 10^{-2}$), $\eta_1$, $\eta_2$ and $\eta_3$ may
increase in the plunging region. However, even in this case, the conditions 
in Eq.~(\ref{nc1}) are always satisfied throughout the plunging region.

\subsection{Self-consistent solutions to the dynamical equations}
\label{sec62}

If we insert the solutions of Maxwell's equations that we obtained in
Sec. \ref{sec4} into Eqs.~(\ref{lcon1}) and (\ref{econ1}), then apply 
Eq.~(\ref{rhm}), we obtain
\begin{eqnarray}
	\frac{d}{dr} \left(F_{\rm m} L + 4\pi r^2\, T_{{\rm EM},
		\phi}^{~~~~~r}\right) &=& 0 \;, 
	\label{dl1} \\
	\frac{d}{dr} \left(F_{\rm m} E - 4\pi r^2\, T_{{\rm EM},
		t}^{~~~~~r}\right) &=& 0 \;,
	\label{de1}       
\end{eqnarray}
where
\begin{eqnarray}
	T_{{\rm EM},\phi}^{~~~~~r} = - \frac{C_0 \Delta}{4 \pi r^4 
		u^r}\left(C_0 u^\phi + \Psi u^t\right) 
		\;, \hspace{1cm}
	T_{{\rm EM},t}^{~~~~~r} = \frac{\Psi}{C_0} 
		T_{{\rm EM},\phi}^{~~~~~r} \;.
	\label{tfr}
\end{eqnarray}
When $\Psi = 0$ [i.e., Eq.~(\ref{pc}) is satisfied in the plunging
region], we have $T_{{\rm EM},t}^{~~~~~r} = 0$. Then Eq.~(\ref{de1})
implies that $E$ keeps constant in the plunging region. However, 
since $T_{{\rm EM}, \phi}^{~~~~~r} \neq 0$, by Eq.~(\ref{dl1}) $L$ 
varies in the plunging region. Therefore, for the solutions satisfying 
the boundary condition (\ref{pc}) [or, equivalently, Eq.~(\ref{psi0})], 
the specific energy of particles is conserved but the specific
angular momentum changes.

Setting $L = u_\phi$, $E = - u_t$, and $\Psi = 0$, Eqs.~(\ref{dl1})
and (\ref{de1}) can be integrated to obtain
\begin{eqnarray}
	u_\phi + \frac{c_0^2 \Delta u^\phi}{r^2 u^r} &=& 
        -f_{\rm L} \;, \label{fl} \\
	u_t &=& f_{\rm E} \;, \label{fe}
\end{eqnarray}
where 
\begin{eqnarray}
	c_0 \equiv \frac{C_0}{\sqrt{-F_{\rm m}}} \;, \hspace{1cm} 
	f_{\rm L} \equiv \frac{F_{\rm L}}{- F_{\rm m}} \;, \hspace{1cm}
	f_{\rm E} \equiv \frac{F_{\rm E}}{- F_{\rm m}} \;,
\end{eqnarray}
where $F_{\rm L} = 4\pi r^2 J_{\rm L}^r$ and $F_{\rm E} = 4\pi r^2 
J_{\rm E}^r$ are constants measuring the angular momentum flux and the 
energy flux across radius $r$, respectively. Using $u^\phi = g^{\phi
\phi} u_\phi + g^{\phi t} u_t$, we can solve Eqs.~(\ref{fl}) and 
(\ref{fe}) for $u_\phi$
\begin{eqnarray}
	u_\phi = \frac{\frac{2 M a c_0^2}{r^3 u^r} f_{\rm E}
		- f_{\rm L}}{1 + \frac{c_0^2}{r^2 u^r} \left(1 - 
		\frac{2 M}{r}\right)} \;.
	\label{uf}
\end{eqnarray}
Then, from $g^{ab} u_a u_b = -1$ and $u_r = g_{rr} u^r$, we obtain the 
equation for $u^r$
\begin{eqnarray}
	\left(r u^r\right)^2 = - \Delta\left(1 - \frac{f_{\rm E}^2}{1- 
		\frac{2 M}{r}}\right)- \frac{1}{1 - \frac{2 M}{r}}
		\left[\frac{\frac{2 M a}{r} f_{\rm E} + \left(1 -
		\frac{2 M}{r}\right) f_{\rm L}}{1 + \frac{c_0^2}{r^2 
		u^r} \left(1 - \frac{2 M}{r}\right)}\right]^2 \;.
	\label{wind}
\end{eqnarray}

Though the factor $\left(1 - \frac{2 M}{r}\right)$ appears in the 
denominators on the right-hand side of Eq.~(\ref{wind}), $r = 2M$ is 
not a singularity of the equation, since the factor $\left(1 - \frac{
2 M}{r}\right)$ disappears from the denominators if we expand the 
second term on the right-hand side of Eq.~(\ref{wind}), then combine 
with the first term. However, the factor $\left[1 +
\frac{c_0^2}{r^2 u^r} \left(1 - \frac{2 M}{r}\right)\right]$ 
represents a singularity of Eq.~(\ref{wind}) at
\begin{eqnarray}
	1 + \frac{c_0^2}{r^2 u^r} \left(1 - \frac{2 M}{r}\right)
		= 0 \;.
	\label{sing1}
\end{eqnarray}

The differentiation of Eq.~(\ref{wind}) with respect to $u^r$ gives rise
to another singularity of the equation, which is at
\begin{eqnarray}
	u_r u^r - \frac{c_0^2}{r^2 u^r} \left(1- \frac{2M}{r} -
		f_{\rm E}^2\right) = 0 \;.
        \label{sing2}
\end{eqnarray}
This singularity appears as one differentiates Eq.~(\ref{wind}) with
respect to $r$ to obtain a differential equation for $u^r$.

If we define the relativistic Alfv\'{e}n velocity by
\begin{eqnarray}
	c_{\rm A}^{~\,a} \equiv \frac{B^a}{\sqrt{4\pi\rho_{\rm m} + 
		B^2}} \;,
\end{eqnarray} 
then we have
\begin{eqnarray}
	1 +\frac{c_0^2}{r^2 u^r} \left(1 - \frac{2 M}{r}
		\right) = \frac{1}{1 - c_{\rm A}^2}
		\left( 1- \frac{c_{{\rm A}r} c_{\rm A}^{~\,r}}
		{u_r u^r}\right) \;,
	\label{crur}
\end{eqnarray}
and
\begin{eqnarray}
	\frac{c_0^2}{r^2 u^r} \left(1- \frac{2M}{r} -
		f_{\rm E}^2\right) = \frac{c_{\rm A}^2}
		{1- c_{\rm A}^2} \;,
	\label{car}
\end{eqnarray}
where $c_{\rm A}^2 \equiv c_{{\rm A}a} c_{\rm A}^{~\,a} <1$. Therefore,
the singularities given by Eqs.~(\ref{sing1}) and (\ref{sing2}) are 
critical points related to the Alfv\'{e}n speed: the Alfv\'{e}n point 
[Eq.~(\ref{sing1})] and the fast critical point [Eq.~(\ref{sing2})]
\cite{web67,lam99}. 

Since we have assumed that the gas pressure is zero, the slow critical 
point is at $u^r = 0$ which is not relevant to us here. The Alfv\'{e}n
point is not an X-type singularity and it does not impose any additional
conditions on the solution for $u^r$ except setting the integral
constant \cite{web67,lam99}. Therefore, what is really relevant here
is the fast critical point given by Eq.~(\ref{sing2}). For weak fields,
we expect that the fast critical point is located at a radius close to
the marginally stable radius, where the accretion flow transits from 
subsonic motion in the disk region to supersonic motion in the plunging 
region \cite{abr89,pac00,afs02}.
Any physical solutions must smoothly pass the critical points, which sets
strict constraints on the integral constants.

It is far beyond the scope of the current paper to fully explore the
properties of critical points in detail. However, as a first order 
approximation, we can assume that $r_{\rm f} \approx r_{\rm ms}$, where 
$r_{\rm f}$ is the radius at the fast critical point. Then, from
Eq.~(\ref{sing2}) we have the radial velocity of the fluid at $r = 
r_{\rm ms}$
\begin{eqnarray}
	u_0^r \approx - \left[\frac{c_0^2 \Delta_{\rm ms}}{r_{\rm ms}^4} 
		\left(-1 +\frac{2M}{r_{\rm ms}} + f_{\rm E}^2 \right)
		\right]^{1/3} \;.
		\label{ur0}
\end{eqnarray}
If we set $r = r_{\rm ms}$ in Eq.~(\ref{wind}), then substitute 
Eq.~(\ref{ur0}) into Eq.~(\ref{wind}), we can solve for the constant 
$f_{\rm L}$, as a 
function of $c_0$ and $f_{\rm E}$. Therefore, for solutions that smoothly 
pass the critical points, among the three integral constants $c_0$, 
$f_{\rm E}$, and $f_{\rm L}$, only two are independent.

Without explicitly solving the radial flow equation, i.e. Eq.~(\ref{wind}), 
we can also obtain some interesting results on the horizon of the black hole.
Since $\Delta\rightarrow 0$ as $r\rightarrow r_{\rm H}$, from 
Eq.~(\ref{wind}) we have
\begin{eqnarray}
	u^r_{\rm H} = \frac{c_0^2 a^2}{r_{\rm H}^4} + \frac{2 M}
		{r_{\rm H}}\left(f_{\rm E} - \Omega_{\rm H} f_{\rm L}
		\right) \;,
	\label{urh}
\end{eqnarray}
where $u_{\rm H}^r = u^r(r = r_{\rm H})$. Then, from Eq.~(\ref{uf}), we
can obtain the specific angular momentum for particles at $r = r_{\rm
H}$
\begin{eqnarray}
	L_{\rm H} = - f_{\rm L} + \frac{a}{r_{\rm H}^2} c_0^2 \;.
\end{eqnarray}

Eq.~(\ref{fe}) says that, with the boundary condition given by 
Eq.~(\ref{psi0}), the specific energy of fluid particles does not change 
in the plunging region. However, the specific angular momentum of fluid
particles does change.
The specific angular momentum for particles on the marginally stable circular
orbit, $L_0$, can be calculated from Eq.~(\ref{uf}) by setting $r = 
r_{\rm ms}$. Then, by comparing $L_{\rm H}$ to $L_0$, we can see how 
much has changed in the specific angular momentum as particles move 
from the marginally stable circular orbit to the black hole horizon. We have 
calculated the ratio 
\begin{eqnarray}
	\frac{\Delta L}{L} \equiv \frac{1}{L_0} \left(L_{\rm H} - 
		L_0\right) \;,
	\label{dell}
\end{eqnarray}
and presented the results in Fig.~\ref{fig12}.

To make the solutions smoothly joined to those in the disk region, in 
our calculations we have adopted that the specific energy is equal to 
that of a particle moving on the marginally stable circular orbit. Then, we 
have $f_{\rm E} = - E_{\rm ms}$. From Eqs.~(\ref{car}) and (\ref{ur0}), 
on the marginally stable circular orbit there is a one-to-one correspondence 
between $c_0^2$ and
\begin{eqnarray}
	\eta_1 = \frac{B^2}{4\pi\rho_{\rm m}}
		 = \frac{c_{\rm A}^2}{1- c_{\rm A}^2} \;.
	\label{eta1}
\end{eqnarray} 
Therefore, we use $\eta_1$ at $r=r_{\rm ms}$ to specify 
the boundary value of the magnetic field, then $c_0$ is determined 
(up to a sign --- which is not relevant here). Then, with the above
approach, $f_{\rm L}$ is determined as a function of $\eta_1$ and 
$f_{\rm E}$. We have chosen $\eta_1$ to be $10^{-2}$, $10^{-3}$, 
and  $10^{-4}$, alternatively.

From Fig.~\ref{fig12} we see that, though $L$ is not constant in the 
plunging region, its variation is extremely small, if on the marginally 
stable circular orbit the dynamical effects of the magnetic field on the
motion of the fluid particles is negligible (then $\eta_1 \ll 1$ at
$r = r_{\rm ms}$). As $a/M$ increases from $-1$ to $1$, the variation in
the specific angular momentum decreases, caused by the fact that the 
coordinate 
distance between the marginally stable circular orbit and the event horizon
decreases with increasing $a/M$. For a black hole with $a/M \ge0$, we 
have $|\Delta L/L| < 3\%$ if $\eta_1 < 10^{-2}$ on $r = r_{\rm ms}$. 
For the extreme case of $a/M = 1$, we have $|\Delta L/L| = 0$, i.e.
the specific angular momentum does not change at all. For the extreme 
case of $a/M = - 1$, which has the largest coordinate distance from the 
marginally stable circular orbit to the event horizon, the variation in 
$L$ is largest. However, even in this case, the variation in the specific
angular momentum is also not big: $|\Delta L/L| < 17\%$ if $\eta_1 < 
10^{-2}$ on $r = r_{\rm ms}$. 

For the same models we have also calculated the ratios $\eta_1$ and 
$c_{{\rm A}r} c_{\rm A}^{~~r}/u_r u^r$ at $r = r_{\rm H}$, and presented 
the results in Fig.~\ref{fig13}. Because of Eq.~(\ref{eta1}), the value 
of $\eta_1$ is given by the left-hand side of Eq.~(\ref{car}). The ratio 
$c_{{\rm A}r} c_{\rm A}^{~~r}/u_r u^r$ can be calculated by
\begin{eqnarray}
	\frac{c_{{\rm A}r} c_{\rm A}^{~~r}}{u_r u^r}
		= \frac{c_0^2 f_{\rm E}^2}{c_0^2 \left(f_{\rm E}^2
		-1 + \frac{2M}{r}\right) - r^2 u^r} \;.
	\label{crr}
\end{eqnarray}
From Fig.~\ref{fig13} we see that both $B^2/4\pi\rho_{\rm m}$ (solid
curves) and $c_{{\rm A}r} c_{\rm A}^{~~r}/u_r u^r$ (dashed curves) are 
small at $r = r_{\rm H}$. For $a/M\ge 0$, both are $< 0.03$ if
$\eta_1 < 10^{-2}$ at $r = r_{\rm ms}$. The ratios go down as $a/M$
increases. Even for the extreme case of $a/M = -1$, the ratios are also 
not big at $r = r_{\rm H}$ if $\eta_1 < 10^{-2}$ at $r = 
r_{\rm ms}$: $B^2/4\pi\rho_{\rm m} < 0.18$ and $c_{{\rm A}r} 
c_{\rm A}^{~~r}/u_r u^r < 0.07$. 

Eq.~(\ref{urh}) implies that $u_{\rm H}^r$ is always finite. Then, 
$u_r u^r = (r^2/\Delta) \left(u^r\right)^2 \rightarrow \infty$ at $r 
= r_{\rm H}$ since $\Delta(r = r_{\rm H}) = 0$. Then, Eq.~(\ref{crr}) 
implies that $c_{{\rm A}r} c_{\rm A}^{~~r}\rightarrow \infty$ also at 
$r = r_{\rm H}$ since its right-hand side is finite. However, from
Eq.~(\ref{crur}) we have
\begin{eqnarray}
	\left.\frac{c_{{\rm A}r} c_{\rm A}^{~\,r}}
		{u_r u^r}\right\vert_{r = r_{\rm H}} < 1
		\;,
	\label{crurh}
\end{eqnarray}
since $1 - 2M/r_{\rm H} = -a^2/r_{\rm H}^2$ and $u_{\rm H}^r < 0$\,.

Since $u_{\rm H}^r$ is finite, Eqs.~(\ref{car}) and (\ref{eta1}) imply 
that $\eta_1$ is also finite at $r = r_{\rm H}$. Then, 
since $u_r u^r\rightarrow\infty$ at $r = r_{\rm H}$, we must have
\begin{eqnarray}
	\left.\frac{c_{\rm A}^2/(1- c_{\rm A}^2)}{u_r u^r}
		\right\vert_{r =r_{\rm H}} = 0 \;.
	\label{crurh2}
\end{eqnarray}
Eqs.~(\ref{crurh}) and (\ref{crurh2}) imply that fluid particles always
supersonically fall into the black hole. The numerical results in 
Fig.~\ref{fig13} confirm this conclusion.

From Figs.~\ref{fig10} -- \ref{fig13}, we can demonstrate that when the 
boundary condition (\ref{pc}) [or, equivalently, (\ref{psi0})] is 
satisfied, the dynamical effects of magnetic fields in the plunging region 
are unimportant if they are so on the marginally stable circular orbit.

\section{Summary and Discussion}
\label{sec7}

With a simple model we have studied the evolution of magnetic fields in
the plunging region around a Kerr black hole. The model contains the
following assumptions: (1) The background spacetime is described by the 
Kerr metric; (2) The plasma is perfectly conducting so that the magnetic 
field is frozen to the plasma fluid; (3) The kinematic approximation 
\cite{bal98}
applies, i.e. the dynamical effects of the magnetic field on the fluid
motion are negligible so that the plasma fluid flows along timelike 
geodesics toward the central black hole; (4) In a small neighborhood of 
the equatorial plane (i.e., $|\pi/2 - \theta| \ll 1$) the magnetic field
and the velocity field have only radial and azimuthal components. The 
assumption (4) implies that $u^\theta = B^\theta = 0$ and $\partial 
u^\theta/\partial \theta = \partial B^\theta /\partial \theta = 0$ on the 
equatorial plane. With above 
assumptions, we have exactly solved Maxwell's equations for axisymmetric
solutions (i.e. $\partial/\partial \phi = 0$). The solutions are 
given by Eqs.~(\ref{fsol1}) and (\ref{fsol2}) [or, equivalently, 
Eqs.~(\ref{br}) and (\ref{bf})], where $C_0$ is a constant measuring
the magnetic flux through a circle in the equatorial 
plane, $\Psi$ is a function determining the time-evolution of the
magnetic field. Both $C_0$ and $\Psi$ are determined by the boundary
conditions. 

The dependence of the solutions on the coordinate time $t$ is given by
the function $\Psi = \Psi (t-\tau)$, where $\tau = \tau(r)$ is defined
by Eq.~(\ref{tau}). The function $\Psi$ is uniquely determined by the 
initial and boundary conditions 
of the magnetic field. The general form of $\Psi$ determines the 
retarded nature of the solutions: at any time the state of the magnetic
field at a radius in the plunging region is determined by the state
of the magnetic field at a larger radius and an earlier time 
(Figs.~\ref{fig1} and \ref{fig2}). This suggests that the stationary 
state of the magnetic field in the plunging region is uniquely 
determined by the boundary conditions at the marginally stable circular orbit. 

\begin{figure}
\vspace{2cm}
\includegraphics[width=15cm]{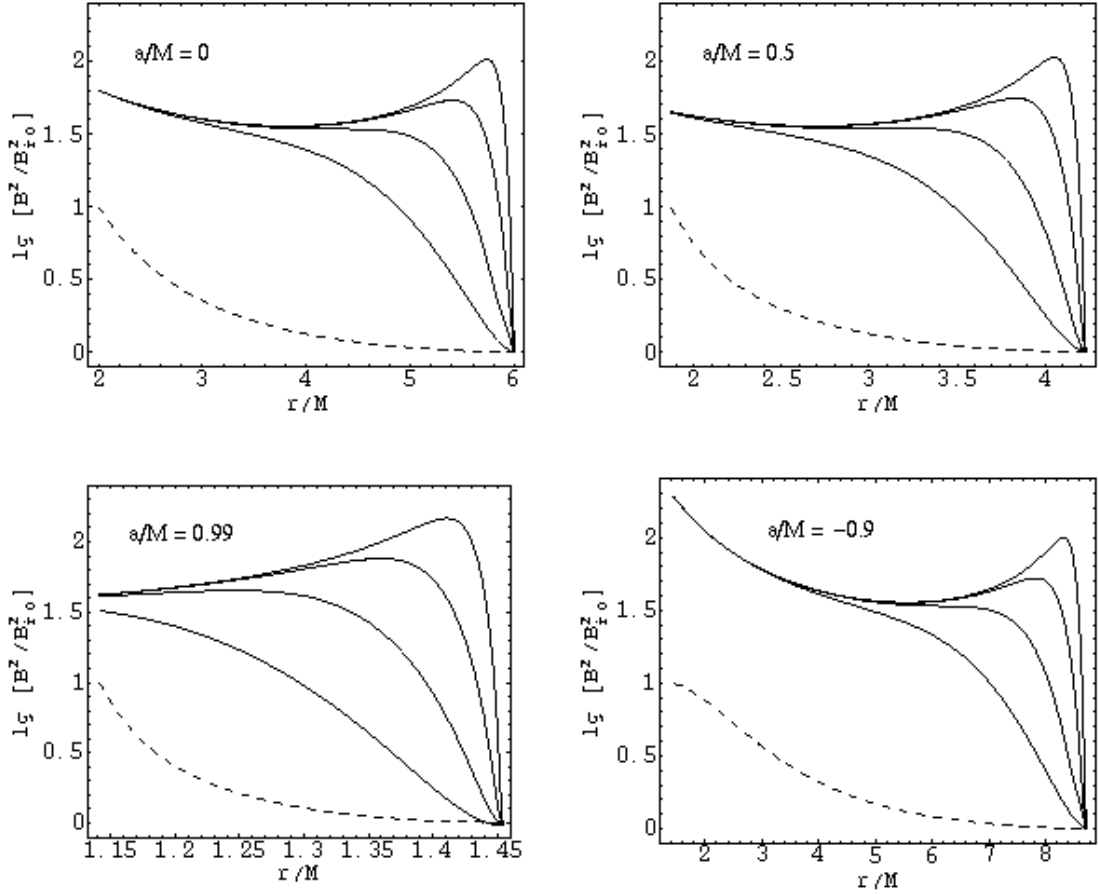}
\caption{\label{fig5} Stationary evolution of $B^2 = B_a B^a$ (solid curves). 
The boundary conditions are the same as that in Fig.~\ref{fig4}. The thin 
dashed line in each panel shows the absolute value of the radial velocity 
(i.e. $|v_{\hat{r}}|$) of the fluid, corresponding to $\delta = 10^{-5}$. 
If we choose a different value of $\delta$, the curve for the radial velocity
will change slightly: 
the right end will change according to $|v_{\hat{r}}|\propto\delta^{1/2}$, 
the left end always approaches $1$ (i.e. the speed of light). (For brevity, 
$\lg \left[B^2 / B_{\hat{r}0}^2\right]$ and $|v_{\hat{r}}|$ use the same 
scale as labeled on the left side of the box in each panel.)}
\end{figure}

\begin{figure}
\vspace{0.5cm}
\includegraphics[width=15cm]{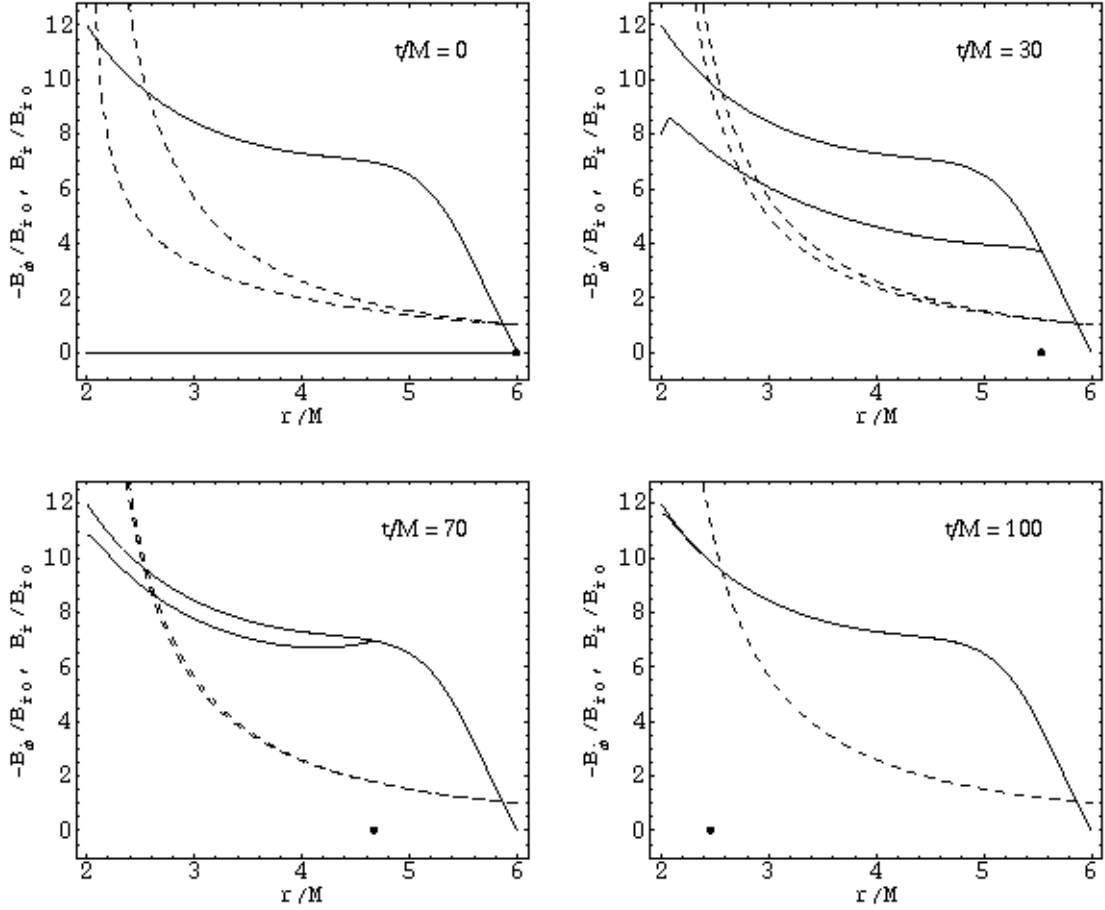}
\caption{\label{fig6} Time evolution of magnetic fields around a 
Schwarzschild black hole (i.e. $a = 0$). Each panel corresponds to a
particular moment. Thick solid lines show the toroidal component
of the magnetic field (i.e., $B_{\hat{\phi}}$), thick dashed lines 
show the radial component (i.e., $B_{\hat{r}}$). The thin (upper) lines show
the corresponding stationary state solutions. The fluid particles  
have specific energy identical to that of the marginally stable circular 
orbit, and specific angular momentum smaller than that of the marginally 
stable circular orbit by a factor of $\delta = 10^{-3}$. The first 
(left and up) panel shows the initial and boundary conditions of the 
magnetic field: on $r = r_{\rm ms}$ (right end), $B_{\hat{\phi}} = 0$, 
$B_{\hat{r}} = B_{\hat{r}0}$ for all time $t>0$; in the plunging
region ($r_{\rm H} < r < r_{\rm ms}$), $B_{\hat{\phi}} = 0$ at $t=0$
[i.e., Eqs. (\ref{bond1}) and (\ref{bond2}); the corresponding 
$B_{\hat{r}}$ at $t = 0$ is automatically determined by the solutions].  
The figures show that, as time goes on, the magnetic field quickly 
approaches and saturates at the state given by the stationary solutions. 
The black dot in each panel shows the radial position of a fluid particle 
at each moment: initially the particle is at $r = r_{\rm ms}$.}
\end{figure}

\begin{figure}
\vspace{4cm}
\includegraphics[width=15cm]{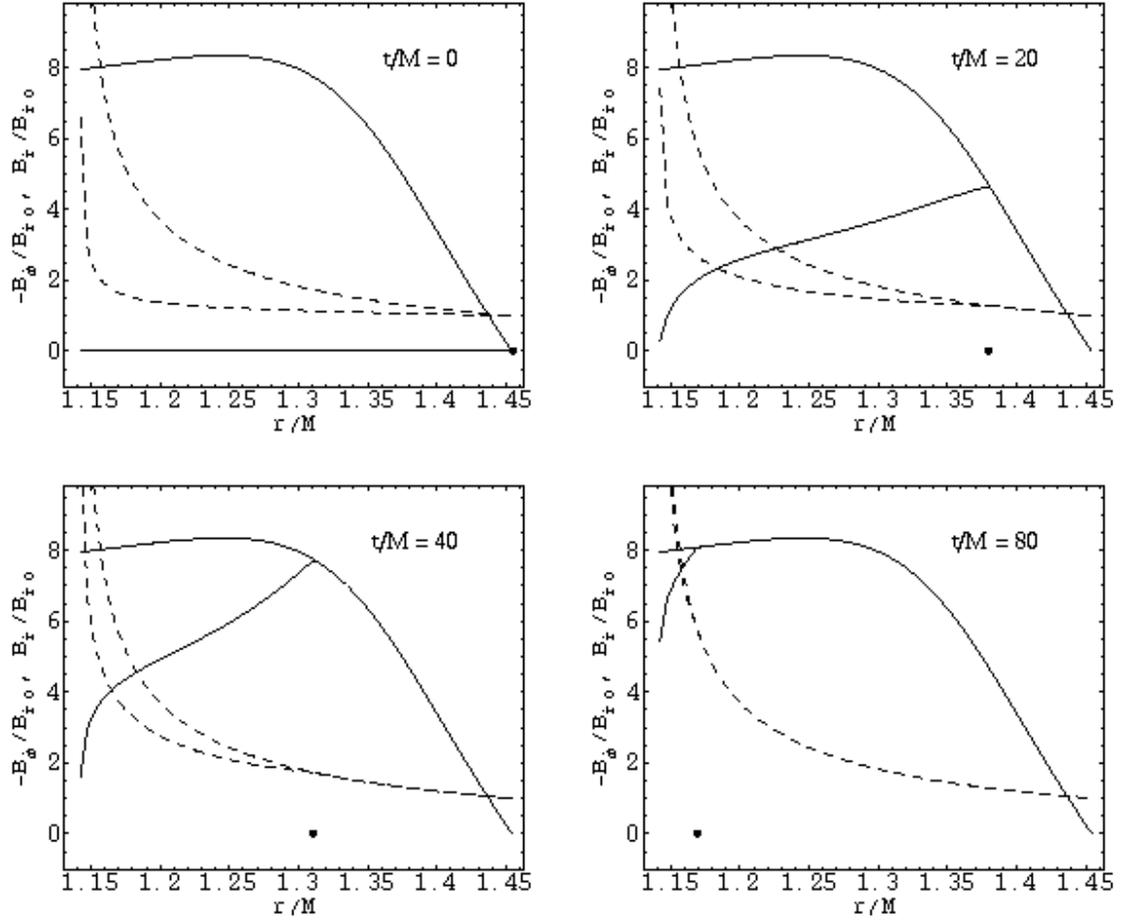}
\caption{\label{fig7} Same as Fig.~\ref{fig6} but for a Kerr
black hole with $a/M = 0.99$.}
\end{figure}

\begin{figure}
\vspace{2cm}
\includegraphics[width=15cm]{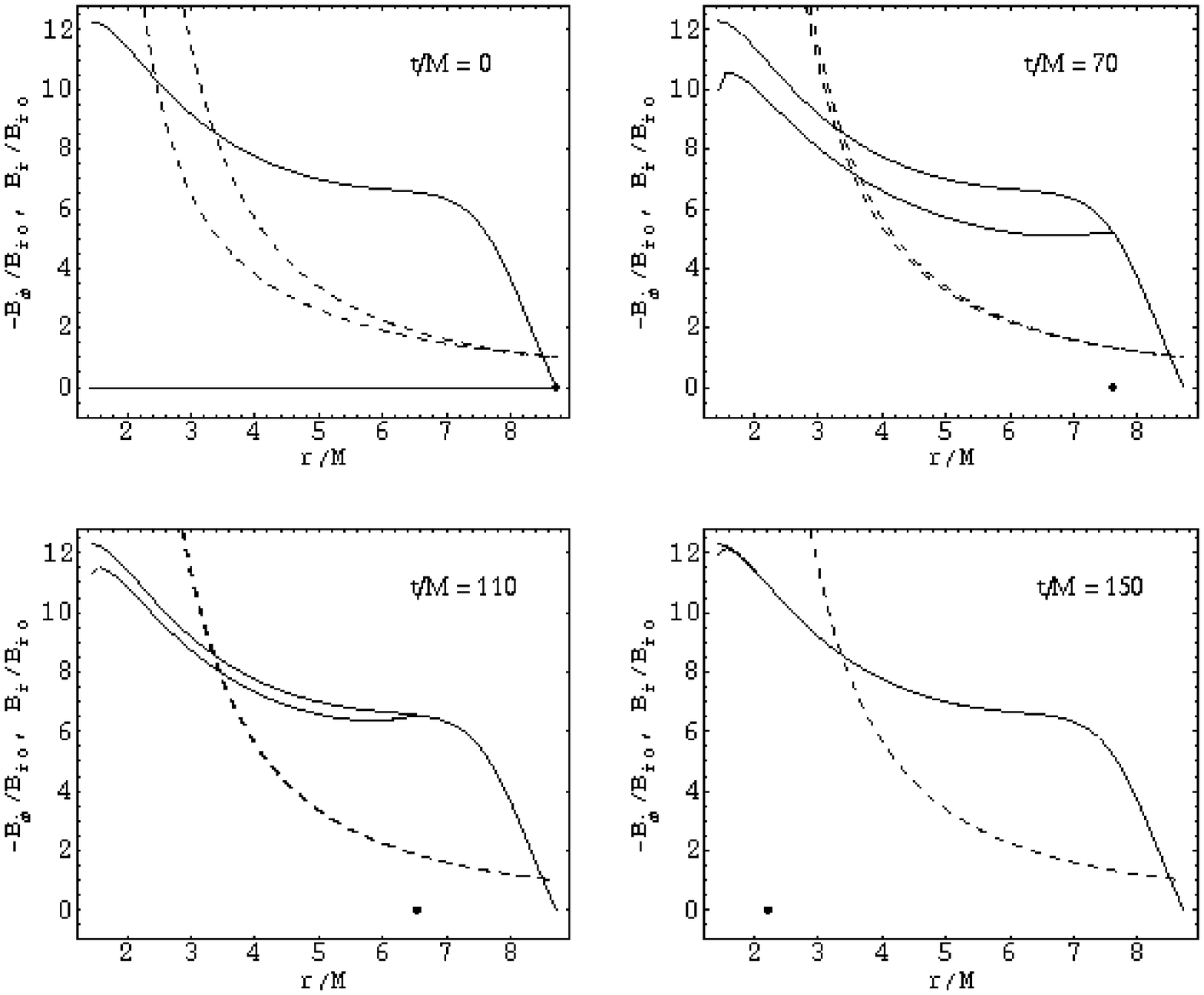}
\caption{\label{fig8} Same as Figs.~\ref{fig6} and \ref{fig7} but 
for a Kerr black hole with $a/M = -0.9$.}
\end{figure}

\begin{figure}
\vspace{2cm}
\includegraphics[width=15cm]{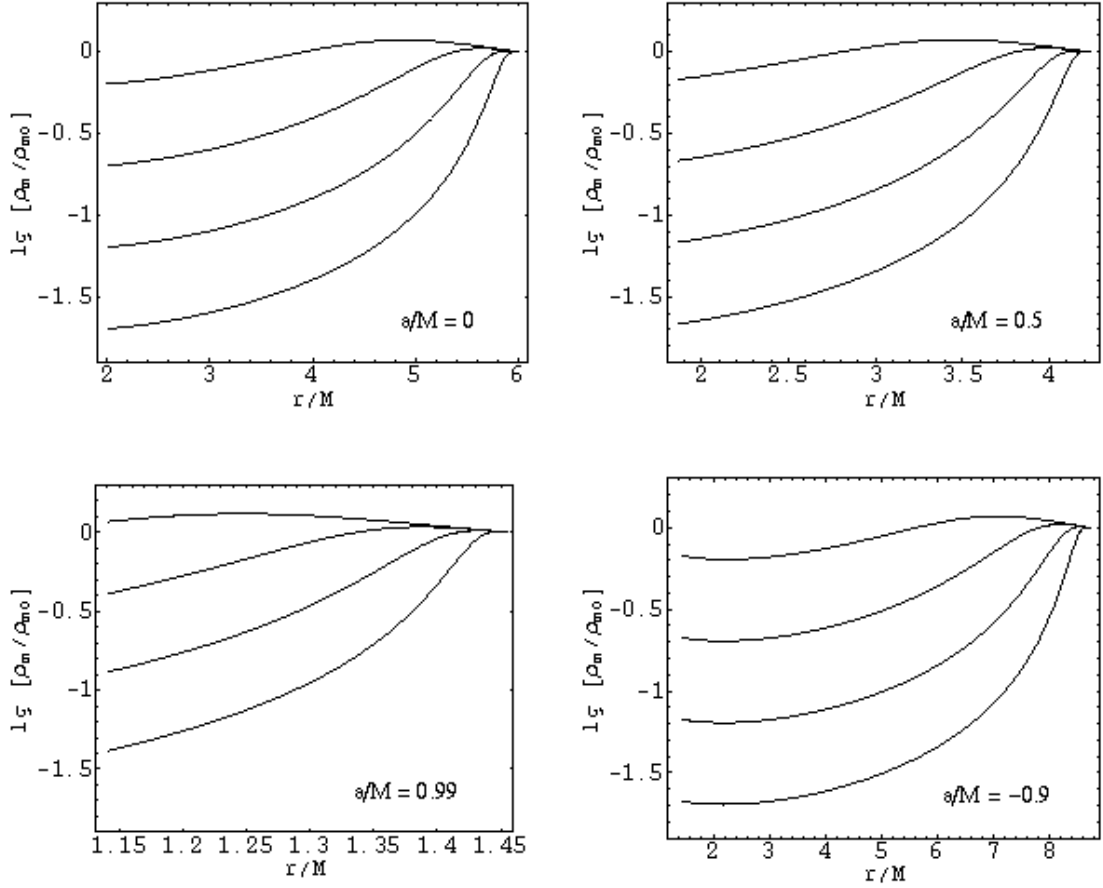}
\caption{\label{fig9} Stationary evolution of the mass density in the 
plunging region around a Kerr black hole. Each panel corresponds to a
different spinning state of the black hole, as indicated by the values
of $a/M$. The right end of each curve corresponds to the marginally
stable circular orbit ($r = r_{\rm sm}$). The left end of each curve 
corresponds to the horizon of the black hole ($r = r_{\rm H}$). The 
kinetic boundary conditions are given by Eq.~(\ref{le0}). In each panel 
the four curves correspond respectively to $\delta = 10^{-2}$, $10^{-3}$, 
$10^{-4}$ and $10^{-5}$ (downward). The mass density is in units of 
$\rho_{{\rm m}0}$ -- the mass density at the marginally stable circular 
orbit.}
\end{figure}

\begin{figure}
\vspace{2cm}
\includegraphics[width=15cm]{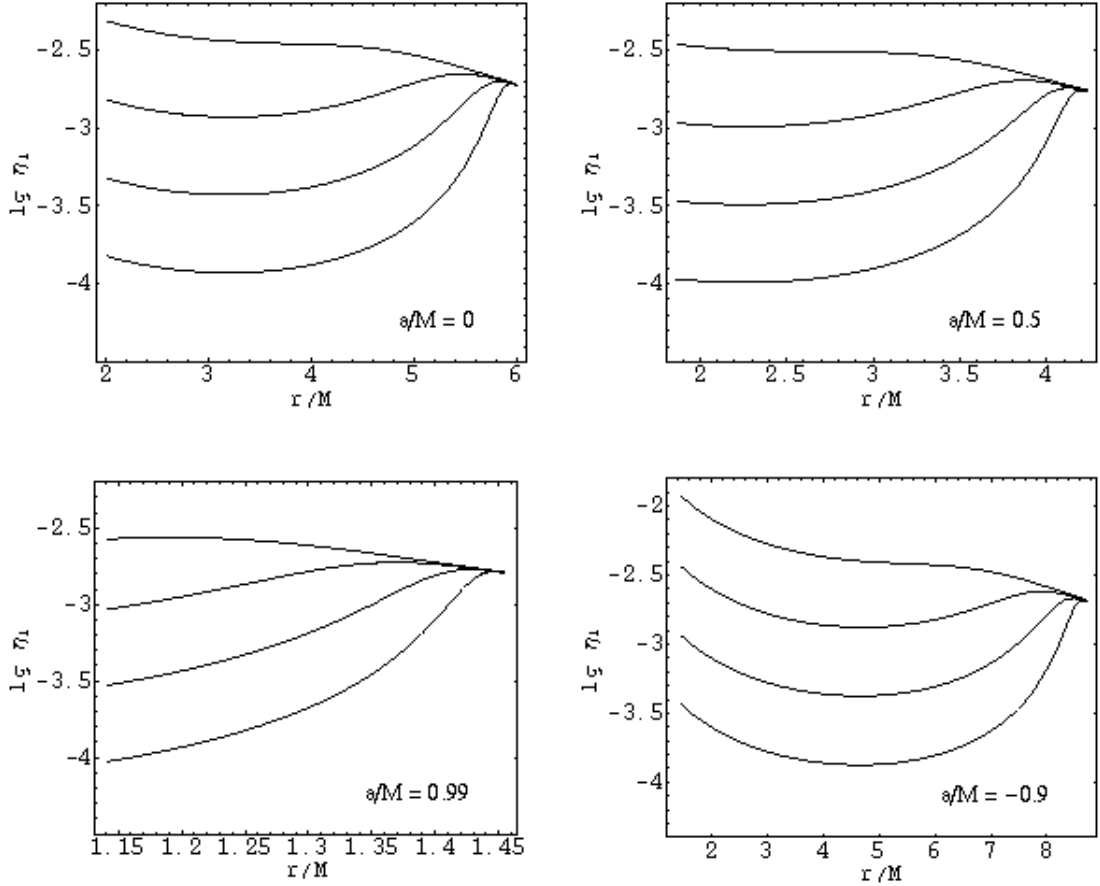}
\caption{\label{fig10} Stationary evolution of the ratio $\eta_1 \equiv 
B^2 /4\pi \rho_{\rm m}$ in the plunging region. Each panel corresponds 
to a different spinning state of the black hole, as indicated by the 
values of $a/M$. The kinetic boundary conditions are given by Eq. 
(\ref{le0}). The boundary conditions for the magnetic field are: on $r 
= r_{\rm ms}$ we have $B_{\hat{\phi}} = 0.05$ in units of $(4\pi \rho_{
{\rm m}0})^{1/2}$. The corresponding $B_{\hat{r}}$ is calculated with
Eq.~(\ref{pc1}) so that the magnetic field is always parallel to the
velocity field. In 
each panel the four curves correspond respectively to $\delta = 10^{-2}$, 
$10^{-3}$, $10^{-4}$ and $10^{-5}$ (downward). Each curve starts from the
marginal stable circular orbit (right end) and ends at the horizon of the
black hole (left end).}
\end{figure}

\begin{figure}
\vspace{4cm}
\includegraphics[width=15cm]{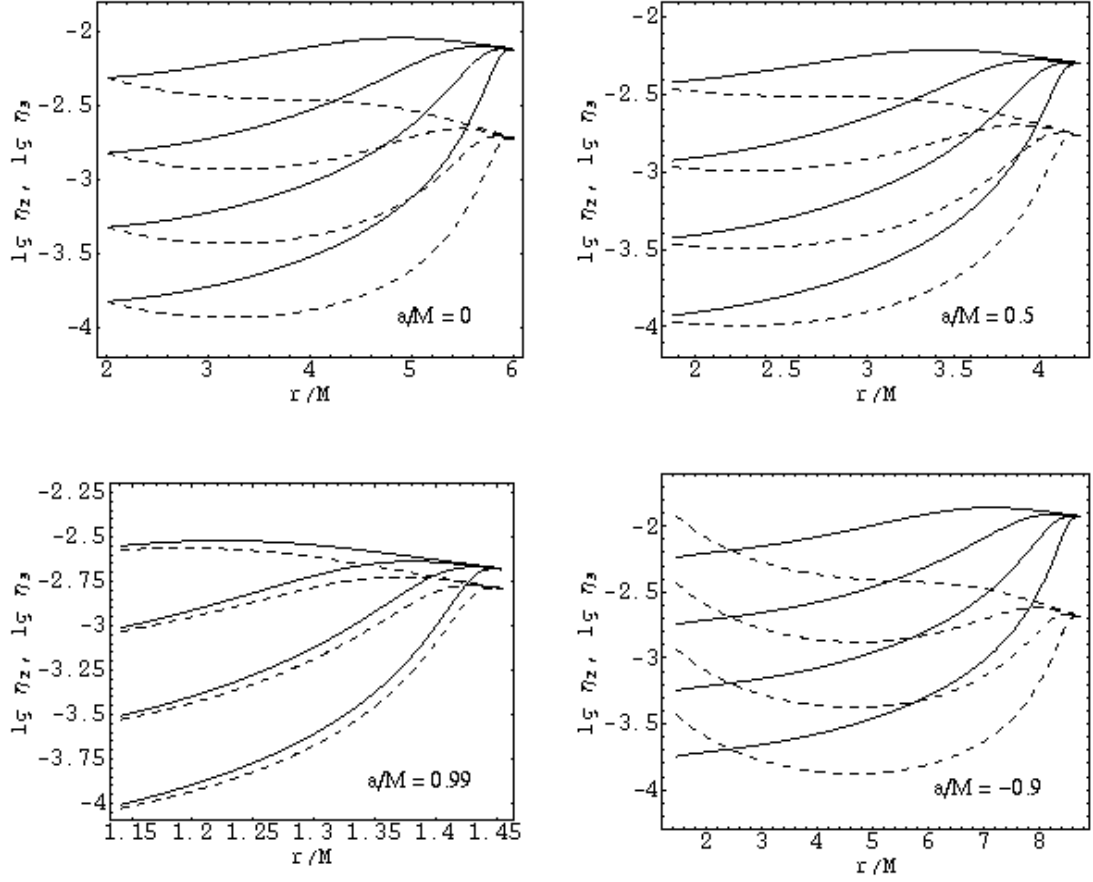}
\caption{\label{fig11} Same as Fig.~\ref{fig10} but for the ratios
$\eta_2 \equiv \left\vert B_\phi B^r / (4\pi\rho_{\rm m} L u^r)
\right\vert$ (solid curves) and $\eta_3 \equiv \left\vert B_t B^r / 
(4\pi\rho_{\rm m} E u^r)\right\vert$ (dashed curves).}
\end{figure}

\begin{figure}
\vspace{3cm}
\includegraphics[width=15cm]{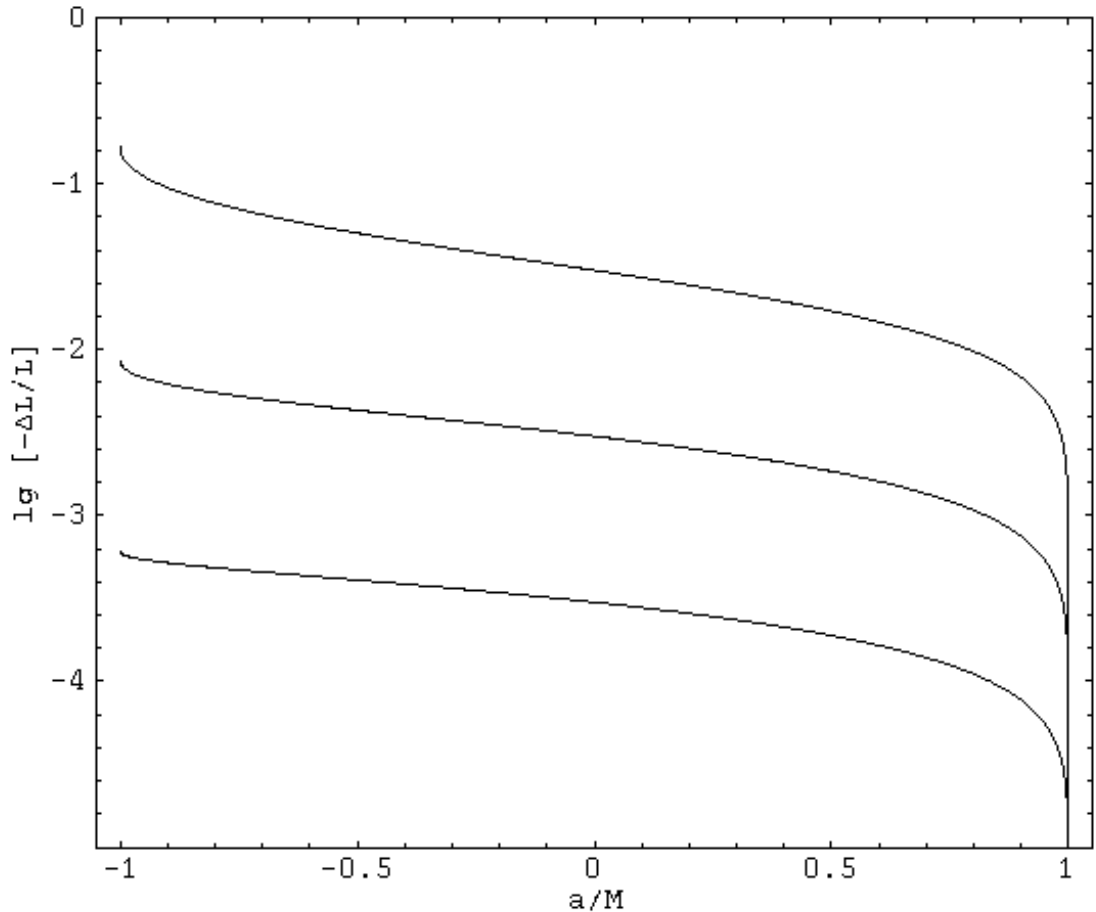}
\caption{\label{fig12} The change in the specific angular momentum 
of fluid particles as they approach the event horizon of the black hole
[Eq.~(\ref{dell})], as a function of the spin of the black hole. It is
assumed that the fast critical point is at $ r \approx r_{\rm 
ms}$\,. The magnetic field is parallel to the velocity field 
of the fluid, then the specific energy of fluid particles keeps constant
in the plunging region, which we have assumed to be equal to the 
specific energy of a particle moving on the marginally stable circular 
orbit. The boundary value of the magnetic field at $r = 
r_{\rm ms}$ is specified by the ratio $\eta_1 = B^2/4\pi\rho_{\rm m}$.
The three curves corresponds to three different values of $\eta_1$
at $r = r_{\rm ms}$\,: $10^{-2}$, $10^{-3}$, and $10^{-4}$ (downward). 
}
\end{figure}

\begin{figure}
\vspace{3cm}
\includegraphics[width=15cm]{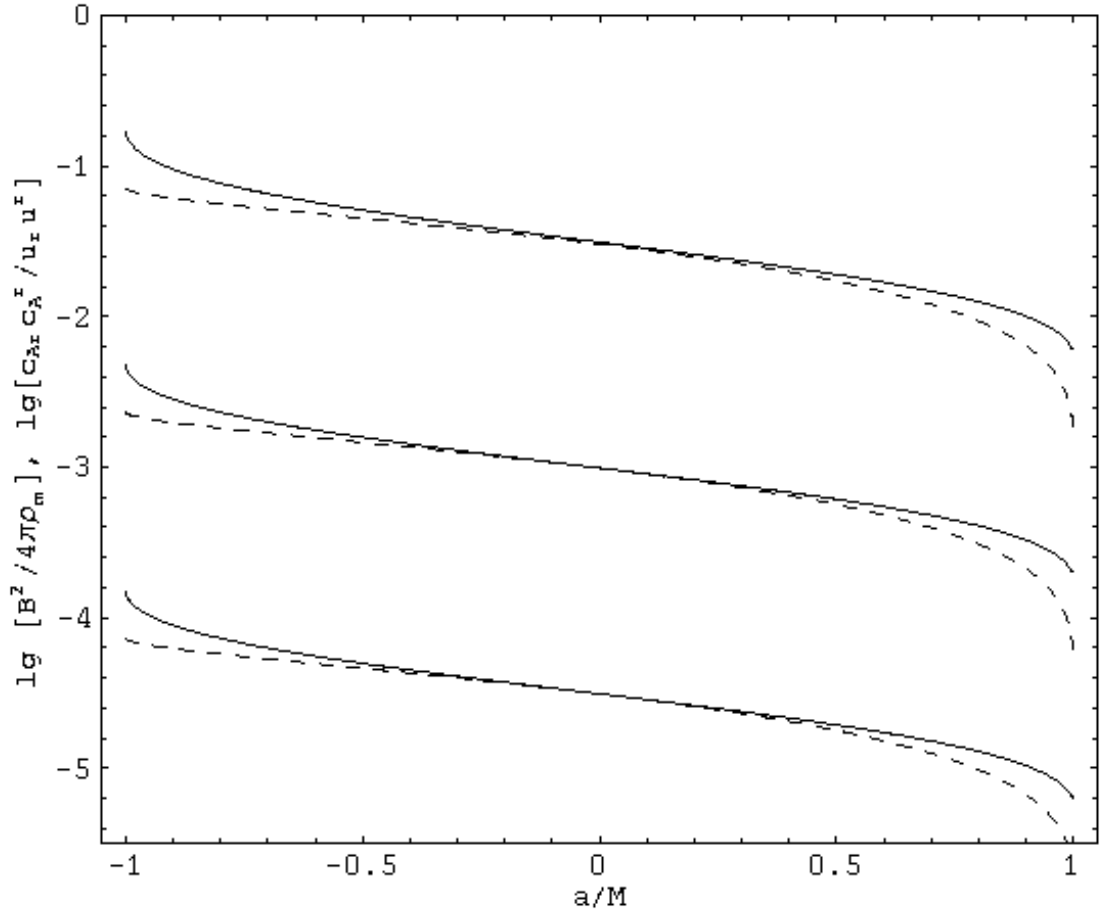}
\caption{\label{fig13} Same as Fig.~\ref{fig12} but for the ratios
$B^2/4\pi\rho_{\rm m}$ (solid curves) and $c_{{\rm A}r} c_{\rm 
A}^{~\,r}/u_r u^r$ (dashed curves) at $r = r_{\rm H}$\,.}
\end{figure}

Examples for the evolution of magnetic fields in the plunging region
are shown in Figs.~\ref{fig3} -- \ref{fig8}, for both the stationary case 
($\Psi = {\rm constant}$, Figs.~\ref{fig3} -- \ref{fig5}) and the 
non-stationary case ($\Psi \neq {\rm constant}$, Figs.~\ref{fig6}
-- \ref{fig8}). From these figures we see that, the boundary value of
the radial component of the magnetic field at the marginally stable circular
orbit is more important in determining the strength of the magnetic field
in the plunging region than the toroidal component. The initially toroidal 
component is attenuated in the plunging region by the radial expansion of 
the fluid (Fig.~\ref{fig3}). The initially radial component is
amplified by the azimuthal and vertical compression (the convergence of 
the fluid), and a toroidal component is generated from the radial component 
by the shear motion of the fluid though the toroidal component is initially 
zero at the marginally stable circular
orbit (Fig.~\ref{fig4}). This leads to the amplification of the magnetic
field in the plunging region (Fig.~\ref{fig5}). The evolution of
the magnetic field depends on the spinning state of the black hole, but
is more sensitive to the initial value of the radial velocity
on the marginally stable circular orbit [or, equivalently the parameter 
$\delta$ defined by Eq.~(\ref{le0})].

The time-evolution of magnetic fields shown in Figs.~\ref{fig6} --
\ref{fig8} confirms our claim that the stationary state of the magnetic
field in the plunging region is uniquely determined by the boundary
conditions on the marginally stable circular orbit. If in the plunging region a 
magnetic field is initially deviated from the stationary solutions, it 
will evolve to and finally get saturated at the state given by the 
stationary solutions in a dynamical time scale determined by the 
free-fall motion of fluid particles in the plunging region. If the 
magnetic field on the marginally stable circular orbit is in a stationary state, 
the magnetic field in the plunging region will automatically settle 
into a stationary state. Thus, the evolution of magnetic fields
in the plunging region is very different from that in the disk region,
in the latter case the state of the magnetic field is determined by
the local MHD processes. The difference is caused by the following fact:
in the disk region the fluid has a very small radial velocity so that
local MHD processes have shorter time scales than the radial motion; 
in the plunging region the fluid has a large radial velocity so that
local MHD processes usually have longer time scales than the radial 
motion.

In deriving the solutions we have assumed that the dynamical effects of 
the magnetic field in the plunging region are unimportant [the kinematic
approximation adopted in assumption 
(3)]. To justify this assumption, we have studied the dynamical 
effects of magnetic fields in the stationary state in two ways. First, 
we estimate the dynamical effects of magnetic fields by considering the 
back-reaction: substituting the solutions of Maxwell's equations
we obtained, where we assumed that fluid particles move on geodesics,
into the dynamical equations to check if the motion of the fluid is 
significantly affected by magnetic fields. In this way, we estimate the
dynamical effects of magnetic fields on the motion of the fluid by
calculating the parameters $\eta_1$, $\eta_2$ and $\eta_3$ defined by 
Eq.~(\ref{eta}). The mass density of the fluid, which appears in the 
denominators of $\eta_1$, $\eta_2$ and $\eta_3$, drops quickly in the 
plunging region if $|v_{\hat{r}}| \ll |v_{\hat{\phi}}|$ on the marginally 
stable circular orbit (Fig.~\ref{fig9}). However, the evolution of the magnetic 
field, which appears in the numerators of $\eta_1$, $\eta_2$ and $\eta_3$,
sensitively depends on the orientation of the magnetic field on the 
marginally stable circular orbit, the latter
is essentially determined by the MHD processes in the disk region. If
we require that the solutions in the plunging region are smoothly joined
to the solutions in the thin Keplerian disk region, then it is 
reasonable to assume that on the marginally stable circular orbit, as well as in
the disk region, the magnetic field is parallel to the velocity field: 
$B^r/B^\phi = u^r/u^\phi$ (see the Appendix). This implies that for the
solutions in the plunging region we should have $\Psi = 0$ 
[Eq.~(\ref{psi0})], i.e. the orientation of the magnetic field 
follows the orientation of the velocity field of the fluid 
[Eq.~(\ref{pc})]. With such a boundary condition, we have calculated 
$\eta_1$, $\eta_2$ and $\eta_3$\,, assuming that particles move on
geodesics. We see that, if they are $\ll 1$ on the marginally stable circular
orbit, then they remain 
so in the plunging region (Figs.~\ref{fig10} and \ref{fig11}). Indeed, 
$\eta_1$, $\eta_2$ and $\eta_3$ decrease in the plunging region for 
sufficiently small $|v_{\hat{r}}|$ on the marginally stable circular orbit.   

Second, we self-consistently solve the coupled Maxwell and dynamical
equations on the horizon of the black hole, check how much has changed
in the specific angular momentum of fluid particles since they leave
the marginally stable circular orbit. The specific energy of fluid particles is 
not changed by a magnetic field that satisfies $\Psi = 0$ [see 
Eq.~(\ref{fe})]. However, the specific angular momentum does change
[see Eq.~(\ref{fl})]. We have calculated the specific angular momentum
of fluid particles as they reach the horizon of the black hole, 
assuming that the fluid particles pass the fast critical point near
the marginally stable circular orbit. The deviation from the specific angular
momentum as the particles just leave the marginally stable circular orbit is
shown in Fig.~\ref{fig12}. We see that, though the specific angular 
momentum of fluid particles is changed by the magnetic field, the effects 
are always small assuming that on the marginally stable circular orbit the 
dynamical effects of the magnetic field are not important. For example,
for $a/M \ge 0$ we have $|\Delta L/L| < 3\%$ if $\eta_1 <10^{-2}$ on
the marginally stable circular orbit. We have also calculated the ratios $\eta_1
= B^2/4\pi\rho_{\rm m}$ and $c_{{\rm A}r} c_{\rm A}^{~\,r}/u_r u^r$ on
the horizon (Fig.~\ref{fig13}), where $c_{\rm A}^{~\,r}$ is the radial
component of the Alfv\'{e}n velocity. These two ratios are relevant
to the critical points in the flow [see Eqs.~(\ref{crur}), (\ref{car}), 
and (\ref{eta1})] and measure at what level the magnetic field affects
the motion of the fluid. Fig.~\ref{fig13} shows that their values on
the horizon are quite small.

Both approaches (back-reaction and self-consistent solutions) confirm 
that the dynamical effects of the magnetic field are unimportant in the 
plunging region if they are so on the marginally stable circular orbit. 

Our results differ from that of Gammie \cite{gam99}, in which he claimed 
that in the plunging region the dynamical effects of magnetic fields can
be important. The difference is caused by the fact that Gammie used
a boundary condition that is different from ours for the magnetic field. 
Gammie assumed that $\Psi = - \Omega_{\rm ms} C_0$, where
$\Omega_{\rm ms}$ is the angular velocity of the disk at the marginally
stable circular orbit, while we assume that $\Psi = 0$. Gammie's boundary
condition implies that $v_{\hat{r}} = 0$ at $r = r_{\rm ms}$ (as clearly
stated in his paper), which makes $r = r_{\rm ms}$ a singular point where 
$\rho_{\rm m} = \infty$ (to keep the mass flux $F_{\rm m}$ nonzero). Hence, 
Gammie's
solutions are not well-behaved at the marginally stable circular orbit. 
Our boundary condition allows the solutions in the plunging region to be 
smoothly joined to the solutions in the disk region (see the Appendix), 
since in our solutions all physical quantities are finite at the marginally
stable orbit. Certainly, to precisely take care of the transition from
the disk region to the plunging region, gas pressure and non-electromagnetic 
stress must be properly taken into account near the inner edge of the
disk.

As mentioned in the Introduction, recently the ``no-torque inner
boundary condition'' for thin accretion disks has been challenged by 
some authors (including Gammie) based on their studies on the 
evolution of magnetic 
fields in the plunging region \cite{kro99,gam99,haw01b,haw02}. The
results in this paper suggest that in the plunging region the
dynamical effects of the magnetic field are not important, if the
solutions in the plunging region are smoothly joined to the the
solutions in the thin Keplerian disk region. Thus, the argument 
against the ``no-torque inner boundary condition'' is not founded.

Finally, we note that in the paper we have neglected dissipative 
processes like magnetic reconnection and ohmic dissipation, and the 
evaporation effect arising from the magnetic buoyancy and MHD 
instabilities. These processes operate in disks to limit the 
amplification of magnetic fields 
\cite[and references therein]{gal79,tou92,bal98}. Certainly it is 
conceivable that they can also operate in the plunging region to 
reduce the amplification effect of magnetic fields. If these 
processes are important, the results of this paper tend to overestimate 
the amplification of magnetic fields in the plunging region. Then,
the dynamical effects of magnetic fields should be weaker than that
we have estimated without considering dissipative processes, which 
will strengthen our conclusion that the dynamical effects of magnetic
fields are unimportant in the plunging region.

\acknowledgments{The author gratefully thanks Ramesh Narayan and the
anonymous referee whose comments and suggestions led to the improvement
of this work. The author also thanks the Institute for Advanced 
Study, Princeton, for hospitality while this work was being done. This 
work was supported by NASA through Chandra Postdoctoral Fellowship 
grant number PF1-20018 awarded by the Chandra X-ray Center, which is 
operated by the Smithsonian Astrophysical Observatory for NASA under 
contract NAS8-39073.}

\appendix*

\section{Magnetic Fields in the Disk Region}
\nonumber

In the disk region ($r > r_{\rm ms}$), viscosity plays an important role
in the dynamics of disk particles. The viscosity transports angular momentum
outward and dissipates energy, which leads to disk accretion. So, in the 
disk region, particles move on non-geodesic worldlines. 

In the vertical direction, the gravity of the black hole is balanced by the 
gradient of the total pressure (= gas pressure + radiation pressure + 
magnetic field pressure) in the disk.
Therefore, in a small neighborhood of the disk central plane, disk particles
are more likely moving on planes parallel to the equatorial plane, rather
than moving radially as in the plunging region. To describe such a motion,
cylindrical coordinates $(t,r,\phi,z)$ are more suitable, where $t$ and
$\phi$ have the same meaning as those in the Boyer-Lindquist coordinates,
to the first order $r$ also has the same meaning as that in the 
Boyer-Lindquist coordinates, but $z \equiv r \left(\frac{\pi}{2} -\theta
\right)$ where $\left\vert \frac{\pi}{2} -\theta\right\vert \ll 1$\,.

In the cylindrical coordinates, in a small neighborhood of the disk central
plane the Kerr metric can be written as \cite{pag74}
\begin{eqnarray}
	ds^2 = -\left(1-\frac{2M}{r}\right) dt^2 -\frac{4M a}
                {r}\, dt d\phi +\frac{r^2}{\Delta}\, dr^2 
		+\frac{A}{r^2}\, d\phi^2 + dz^2 \;,
    \label{gab2}
\end{eqnarray}
where $\Delta$ and $A$ are defined by Eq.~(\ref{del}) with $\theta =
\pi/2$\,.

The Maxwell equations that we need to solve are again given by 
Eq.~(\ref{maxeq}), but now we have $x^\alpha = (t,r,\phi,z)$ and $\sqrt{-g} 
= r$. As in the case for the plunging region, we assume that the motion of 
fluid particles is stationary but the evolution of the magnetic field can 
depend on time, i.e., we let $\partial u^{\alpha}/\partial t = 0$ but keep 
$\partial B^{\alpha}/\partial t$ in the Maxwell equations. Furthermore, we
assume that in a small neighborhood of the disk central plane $u^z = B^z 
=0$ so that $\partial u^z/\partial z = \partial B^z/\partial z = 0$ on the 
equatorial plane. Finally, we adopt $\partial u^\alpha/\partial\phi = 
\partial B^\alpha/\partial\phi = 0$ because of the axisymmetry of the 
system. Then, on the equatorial plane Eq.~(\ref{maxeq}) is reduced to
\begin{eqnarray}
	\frac{\partial}{\partial t} \left[r (u^t B^\beta - 
		u^\beta B^t)\right] +
		\frac{\partial}{\partial r} \left[r (u^r B^\beta 
		- u^\beta B^r)\right] = 0 \;.
	\label{eq1a}
\end{eqnarray}

Eq.~(\ref{eq1a}) can be solved with the same approach as that used in 
Sec. \ref{sec4}. The solutions are
\begin{subequations}
\begin{eqnarray}
	B^r &=& \frac{1}{r} (-C_{0,{\rm d}} u_t + \Psi_{\rm d} u_\phi) \;, 
		\label{fsol1d}\\
	B^\phi &=& \frac{1}{r u^r} [-C_{0,{\rm d}} u_t u^\phi + (1+ 
                u^\phi u_\phi)\Psi_{\rm d}] \;,
	\label{fsol2d}
\end{eqnarray}
\end{subequations}
where $C_{0,{\rm d}}$ is an integral constant, $\Psi_{\rm d} = \Psi_{\rm d} 
(t-\tau)$ is the solution of Eq.~(\ref{eq4}). The corresponding $B^t$ is
related to $B^r$ and $B^\phi$ by Eq.~(\ref{cons}).

Eq.~(\ref{fsol1d}) can be written as
\begin{eqnarray}
        \Psi_{\rm d} = \frac{1}{u_\phi}(r B^r + C_{0,{\rm d}} u_t) \;.
        \label{psid}
\end{eqnarray}
Assume that as $r\rightarrow \infty$ the disk is Keplerian so that $u_\phi
= L \propto r^{1/2}$ and $u_t = -E \approx 1$, and $r B^r$ keeps finite,
then we must have 
\begin{eqnarray}
        \lim_{r\rightarrow\infty}\Psi_{\rm d} = 0 \;.
	\label{limp}
\end{eqnarray}
Eq.~(\ref{limp}) has an important implication for stationary solutions. For 
stationary solutions we have $\Psi_{\rm d} = {\rm constant}$, then by 
Eq.~(\ref{limp}) we must have $\Psi_{\rm d} = 0$. So, for stationary 
solutions we have
\begin{subequations}
\begin{eqnarray}
	B^r &=& \frac{C_{0,{\rm d}} E}{r} \;, \label{fsol3d}\\
	B^\phi &=& \frac{C_{0,{\rm d}} E u^\phi}{r u^r} \;,
	\label{fsol4d}
\end{eqnarray}
\end{subequations}
where we have used $u_t = -E$. 

From Eqs.~(\ref{fsol3d}) and (\ref{fsol4d}) we have $B^\phi/B^r = 
u^\phi/u^r$, i.e. in the stationary state the magnetic field lines
are parallel to the fluid motion.


\clearpage

\clearpage

\clearpage

\clearpage

\clearpage

\clearpage

\clearpage

\clearpage

\clearpage

\clearpage

\clearpage

\clearpage

\clearpage

\end{document}